\documentclass[useAMS,usenatbib]{mn2e}
\usepackage{epsfig,plotmore,aastex_hack,lscape,amsmath}

\def\lsim{\mathrel{\hbox{\rlap{\hbox{\lower4pt\hbox{$\sim$}}}\hbox{$<$}}}}
\def\gsim{\mathrel{\hbox{\rlap{\hbox{\lower4pt\hbox{$\sim$}}}\hbox{$>$}}}}

\def\LCDM{$\Lambda$CDM }

\newcommand{\kpch}{\>h^{-1} {\rm kpc}}

\title[Shape of Dark Matter Halos]
{The Shape of the Gravitational Potential in Cold Dark Matter Halos}

\author[E.~Hayashi et al.]
{Eric Hayashi$^1$, Julio F.~Navarro$^{2,3}$ and Volker Springel$^1$\\
$^1$Max-Planck Institute for Astrophysics, Garching, Munich, D-85740, Germany \\
$^2$Department of Physics and Astronomy, University of Victoria, Victoria, BC,
V8P 1A1, Canada \\
$^3$Fellow of the Canadian Institute for Advanced Research and of the
J.S.Guggenheim Memorial Foundation}
\begin{document}
\maketitle

\begin{abstract}
We use a set of cosmological N-body simulations to investigate the
structural shape of galaxy-sized cold dark matter (CDM) halos.  Unlike
most previous work on the subject---which dealt with shapes as
measured by the inertia tensor---we focus here on the shape of the
gravitational potential, a quantity more directly relevant to
comparison with observational probes. A further advantage is that the
potential is less sensitive to the effects of substructure and, as a
consequence, the isopotential surfaces are typically smooth and well
approximated by concentric ellipsoids. Our main result is that the
asphericity of the potential increases rapidly towards the centre of
the halo. The radial trend is more pronounced than expected from
constant flattening in the mass distribution, and reflects a strong
tendency for dark matter halos to become increasingly aspherical
inwards. Near the centre the halo potential is approximately prolate
on average ($(c/a)_0 = 0.72 \pm 0.04$, $(b/a)_0=0.78 \pm 0.08$), but
it becomes less axisymmetric and more spherical in the outer
regions. The principal axes of the isopotential surfaces remain well
aligned, and in most halos the angular momentum tends to be parallel
to the minor axis and perpendicular to the major axis.  This suggests
that galactic disks may form in a plane where the potential is
elliptical and where its ellipticity varies rapidly with radius. This
can result in significant deviations from circular motion in systems
such as low surface brightness galaxies (LSBs), even for relatively
minor deviations from circular symmetry. Simulated long-slit rotation
curves often appear similar to those of LSBs cited as evidence for
constant density ``cores''. This suggests that taking into account the
3D shape of the dark mass distribution might help to reconcile such
evidence with the cuspy mass profile of CDM halos.
\end{abstract}


\section{Introduction}
\label{sec:intro}

The flat rotation curves of disk galaxies \citep{RUBIN70,ROBERTS75},
together with the anomalously high velocity dispersion of galaxies in
clusters \citep{ZWICKY33}, constituted the first compelling evidence
for the existence of a massive dark matter component in the Universe.
Since the interpretation of these data depends on the structure of
dark matter halos, many theoretical studies have focused on the radial
distribution of mass within virialized dark matter halos, both
analytically \citep{GUNN72, FILLMORE84, HOFFMAN85} and numerically
\citep{FRENK85,FRENK88,QUINN86,DUB91,CRONE94}.  This effort culminated
in the realization that, in the prevailing cold dark matter (CDM)
paradigm, the spherically averaged mass profile of dark halos is
accurately described by scaling an approximately ``universal'' profile
\citep[][NFW]{NFW96,NFW97}.  The fitting formula proposed by NFW has
enabled a simple means of testing the theoretical expectations against
observations. Overall, these studies have shown that the mass profile
of cold dark matter halos is broadly consistent with observations of
X-ray clusters, gravitational lensing, satellite dynamics, and galaxy
group properties
\citep{POINTECOUTEAU05,COMERFORD06,PRADA03,MANDELBAUM06B}, although
problems remain, especially on the scale of galaxies.

A particularly contentious debate has centred on the inner slope of
the density profile inferred from rotation curves of low surface
brightness (LSB) disk galaxies.  Many authors have argued for the
presence of a constant density `core' at the centre of galactic dark
matter halos, in direct contradiction with the `cuspy', centrally
divergent density profile of simulated CDM halos \citep[see,
e.g.,][]{FLORES94,MOORE94,MCGAUGH98,DEBLOK01}. This conclusion relies
on the assumption that the disk is in circular motion and, therefore,
that the observed rotation velocity is a fair tracer of the circular
velocity of the halo. This assumption holds only for thin, cold,
gaseous disks in spherically symmetric potentials and, while this is a
convenient assumption for preliminary studies, it has long been known
that CDM halos are not spherically symmetric objects
\citep{FRENK88,DUB91,WARREN92,THOMAS98,JING02,BAILIN05, BETT06}.

Recently, \citet[][hereafter HN]{HAYASHI06} investigated the
kinematics of a gaseous disk in a triaxial halo potential using
analytic solutions for closed loop orbits.  The shape of the disk
rotation curve, as measured by a long-slit spectrum, depends on the
orientation of the slit, the inclination of the disk, and the radial
dependence of the departures from spherical symmetry.  HN show that
significant deviations from circular motion may result even in very
mildly triaxial potentials, since the orbital shape is controlled by
the ratio of escape to circular velocities, a quantity that increases
steadily inwards in CDM halos.

Few signatures of the halo triaxial structure---which may be
recognised in 2D velocity fields---are discernible in long-slit
observations. Linearly-rising ``core-like' rotation curves may occur
if (i) the slit samples velocities near the long axis of the
(elliptical) closed orbits and if (ii) the deviations from spherical
symmetry in the halo potential become more pronounced towards the
centre.  As noted above, the shapes of CDM halos have been studied
previously; however, these studies have focused on inertia-tensor
shapes rather than on the potential, thus it is still unclear whether
the radial dependence of halo triaxiality is quantitatively consistent
with the latter requirement.

Establishing the typical radial dependence of triaxiality in the
potential of CDM halos is therefore one of the main concerns of this
paper. The interest of this study goes beyond the topic of LSB
rotation curves, since the non-spherical nature of CDM halos affects
the interpretation of a variety of observational probes, including the
kinematics of polar ring galaxies \citep{SACKETT94}, the warping and
flaring of galactic disks \citep{TEUBEN91, OLLING00}, and the
morphology of tidal streams.  In a spherical potential, tidal streams
remain confined to a plane, whereas the plane of motion will precess
gradually in non-spherical halos. \cite{IBATA01} use this to interpret
tidal debris from the Sagittarius dwarf galaxy, and conclude that the
halo of the Milky Way is spherical and therefore possibly in conflict
with CDM predictions. However, \cite{HELMI04} argues that the
Sagittarius stream may be too dynamically young to be sensitive to the
shape of the dark matter halo, and hence may be consistent with a
Milky Way halo as triaxial as a typical simulated CDM halo \citep[see
also][]{LAW05, JOHNSTON05}.

In galaxy clusters, the hot intracluster gas is in equilibrium with the
gravitational potential of the cluster, and the isodensity surfaces of the gas
coincide with the isopotential surfaces of the cluster halo \citep{BUOTE94,
BUOTE02, LEE03, FLORES05}.  Observations of X-ray emission and
Sunyaev-Zel'dovich (S-Z) decrement measure the integrated density along the
line-of-sight though the cluster, therefore the shape of the X-ray isophotes and
S-Z contours reflects the shape and orientation of the underlying dark matter
halo potential.  Models of the projected surface brightness profiles as a
function of cluster orientation and triaxiality have been developed by
\cite{LEE04} and \cite{WANG04}, but these are based on a constant triaxiality in
the gas isodensity surfaces, which may not be an accurate representation of
the distribution of gas in equilibrium with a realistic CDM halo potential  

There is therefore considerable interest in firming up the theoretical
expectation for the shape and radial dependence of the gravitational
potential in CDM halos, and this paper aims to provide guidance for
such modelling. We focus here on the shape of the isopotential surfaces
of galaxy-sized dark matter halos using high-resolution cosmological
N-body simulations.

The outline of the paper is as follows.  In Section~\ref{sec:sims} we
describe the simulations and methods we use to calculate halo shapes.
In Section~\ref{sec:shapes} we present our measurements of the shapes of
halos and of the internal alignment of halo shapes.  In
Section~\ref{sec:model} we investigate the flattening of the mass
distribution required to reproduce the radial dependence of the halo
potential shapes, and we present a simple fitting formula to
approximate the potential of a triaxial CDM halo.  Finally, we apply
this result in Section~\ref{sec:disk} to the kinematics of disks in
realistic triaxial CDM halo potentials.  We conclude with a brief
summary in Section~\ref{sec:concl}.

\section{Simulations and Methods}
\label{sec:sims}

This study is based on a suite of cosmological N-body simulations of
seven Milky Way-sized galaxy halos and four dwarf galaxy-sized halos.
The concordance \LCDM cosmology is adopted for these simulations, with
$\Omega_0=0.3$, $\Omega_\Lambda=0.7$, $\sigma_8=0.9$. and either
$h=0.65$ (runs labelled G1, G2 and G3) or $h=0.7$ (the rest). These
simulations make use of the techniques described in detail in
\cite{POWER03} and \cite{NAVARRO04} for resimulating halos at high
resolution using nested regions of particles with different mass
resolution and gravitational softening lengths in order to maximize
the number of particles which end up in the target halo at $z=0$.

Each halo contains about $N\simeq 10^6$ particles within the virial
radius, $r_{200}$, defined as the radius of a sphere of mean density
equal to $200$ times the critical density for closure. The
spherically-averaged mass profile of these halos is robust down to
$r_{\rm conv} \simeq 0.01~r_{200}$ according to the convergence
criteria described in \cite{POWER03}.  Table~\ref{tab:halos}
summaries the main properties of the simulated halos; a full
description of the numerical details of these simulations and an
analysis of the halo mass profiles is presented in \cite{HAYASHI04}
and \cite{NAVARRO04}.
%%%% NOTE: values in tab:halos are corrected from Table 3 of Navrro et al 

\begin{figure}
\plotfour{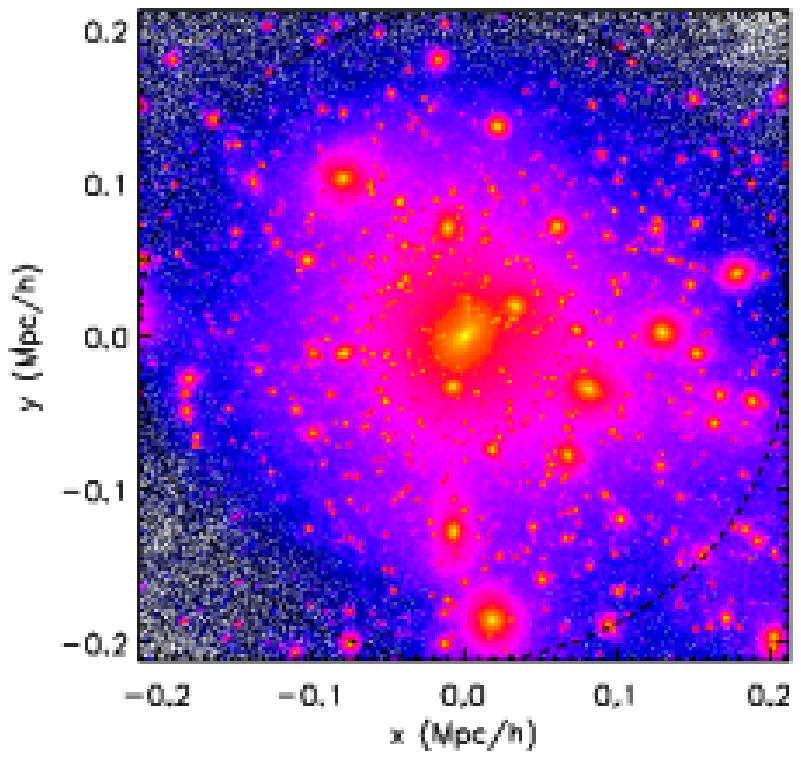}{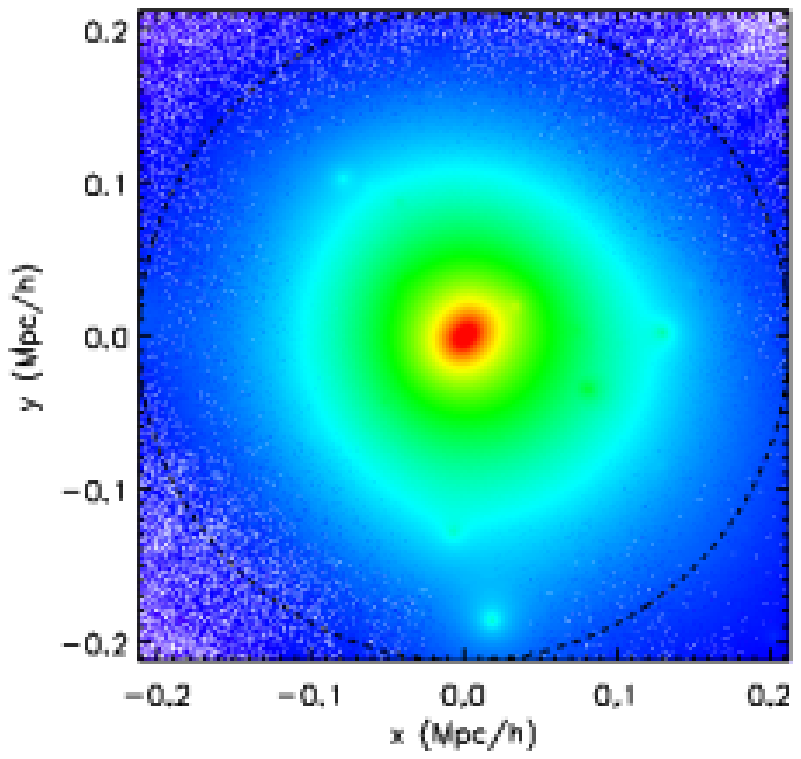}{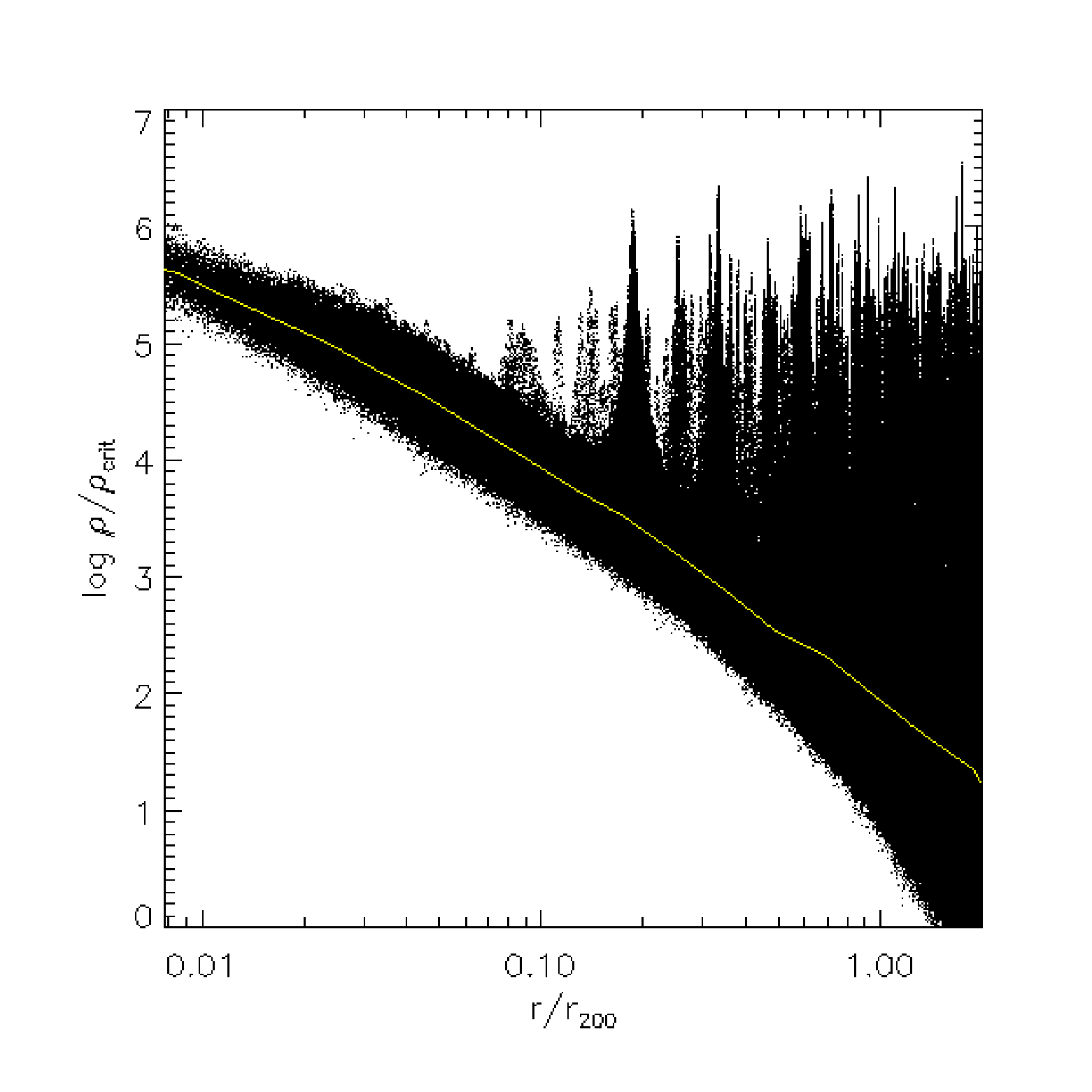}{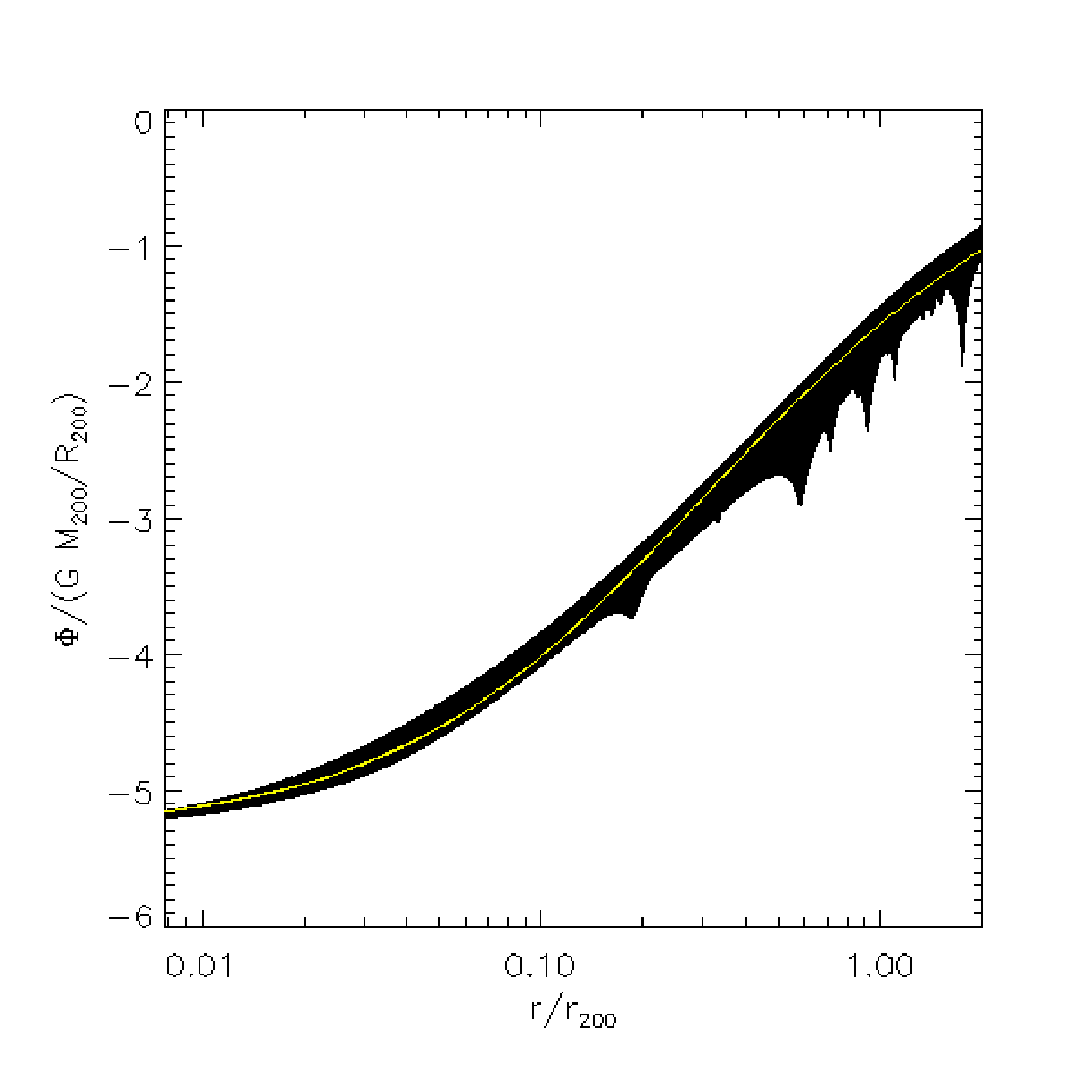}
\caption{ {\it Upper panels:} Halo G3/$256^3$ coloured by local density
(left) and potential (right). Color scheme is logarithmic for density
and linear for the potential.  The virial radius, $r_{200}$, is shown
by the dashed circle.  {\it Lower panels:} Local density (left) and
potential (right) of particles as a function of radius.  Thin solid
lines show the average local density (left panel) and best fit NFW
potential (right panel).
\label{fig:pospotplt}}
\end{figure}

Figure~\ref{fig:pospotplt} shows one galaxy-sized halo from our set of
simulations.  The upper panels show the halo with its particles
coloured according to the values of their local density (left) and
gravitational potential (right).  The local density of each particle
is computed by averaging over its 64 nearest neighbour particles with
a spline kernel similar to that used in Smoothed Particle
Hydrodynamics (SPH) calculations.\footnote{See {\tt
http://www-hpcc.astro.washington.edu/tools/smooth.html}} The
gravitational potential of each particle is computed using all
particles within $4~r_{200}$ of the halo centre, determined using an
iterative technique in which the centre of mass of particles within a
shrinking sphere is computed recursively until a few thousand
particles are left \citep{POWER03}.  The bottom panels plot the values
of the density and gravitational potential of each particle as a
function of its distance from the halo centre.

The most striking feature of this figure is that the wealth of halo
substructure readily apparent in local density, is much less prevalent
in the gravitational potential.  This is not only because the
potential is an integral (and therefore more uniform) quantity, but
also because the total amount of mass associated with substructure is
typically only $5-10\%$ of the mass within $r_{200}$ \citep[see,
e.g.,][]{DELUCIA04}. The lower left panel of
Figure~\ref{fig:pospotplt} shows that particles within substructures
(subhalos) appear as sharply defined ``spikes'' which can exceed the
mean local density at the radius of the subhalo by many orders of
magnitude.  This can cause undesirable biases when measuring halo
shapes.  For instance, \cite{JING02} estimate shapes using isodensity
surfaces, a procedure that requires a careful, and somewhat arbitrary,
removal of substructure as a preliminary step, and that may thus
introduce ambiguities in the results.

In comparison, as emphasised by \cite{SPRINGEL04}, the isopotential
surfaces are much smoother and relatively insensitive to the presence
of substructure.  The lower right panel of Figure~\ref{fig:pospotplt}
shows that the gravitational potential of particles in subhalos
deviates from the mean value at the subhalo radius by at most a factor
of two.  Furthermore, it is the gravitational potential and not the
local density that is the more relevant quantity in most dynamical
studies of halo structure.

In order to measure the three-dimensional shape of a halo's isopotential
surfaces, we adopt the method described by \cite{SPRINGEL04}.  The gravitational
potential is first computed on three uniform grids covering orthogonal planes
which intersect at the location of the minimum gravitational potential of the
halo. In our standard set-up, these meshes have $1024^2$ cells and extend to a
distance of $2~r_{200}$.  We use a hierarchical tree algorithm to compute the
potential efficiently and include all the mass out to a radius of $4~r_{200}$
when processing an individual halo, while more distant particles are ignored.

If the isopotential surfaces are ellipsoids, then their intersections with the
three principal planes are ellipses, and the full shape information of each
ellipsoid can be recovered uniquely from the shapes of the intersections.
Compared to a full 3D grid, using just three planes has the important
advantage of allowing a much finer mesh while reducing the memory and
computational requirements by several orders of magnitudes.

In our actual fitting procedure, we first determine in each plane the
intersections of 360 radial rays with constant angular separation with the
isopotential ellipses for a chosen value of the potential. We define the
potential at every point in the planes by bi-linear interpolation in the
corresponding two-dimensional grid. We adopt the centre-of-mass of the
resulting set of points as the centre of the isopotential ellipsoid, and
identify its principal axes with the eigenvectors of the moment-of-inertia
tensor of the sample of points. To determine the axis lengths $(a, b, c)$ , we
first express the coordinates of the points in the orthonormal basis of the
principal axis relative to the ellipsoid's centre.  Denoting these coordinates
with $(x_i, y_i, z_i)$ we then define a normalized radius $r_i$ for each point
as 
\begin{equation} 
\frac{x_i^2 + y_i^2 + z_i^2}{r_i^2} = \frac{x_i^2}{a^2} +
  \frac{y_i^2}{b^2} + \frac{z_i^2}{c^2}. \label{eqdist} 
\end{equation}
If a point lies on the ellipsoid, $r_i$ would just equal the Cartesian
distance $\sqrt{x_i^2+y_i^2+z_i^2}$, and the left-hand-side of equation
(\ref{eqdist}) would be unity. To determine the best-fitting axis lengths, we
therefore minimize the quantity
\begin{equation}
S= \sum_i \left( r_i -
  \sqrt{x_i^2+y_i^2+z_i^2}\right)^2 
\end{equation} 
with respect to $(a, b, c)$. In practice, we use a Newton-Raphson method to
alternatingly find the roots in $\frac{\partial S}{\partial a} = 0$,
$\frac{\partial S}{\partial b} = 0$, and $\frac{\partial S}{\partial c} = 0$,
cycling through the equations until no further reduction of $S$ can be
achieved.

\begin{figure}
\plotsix{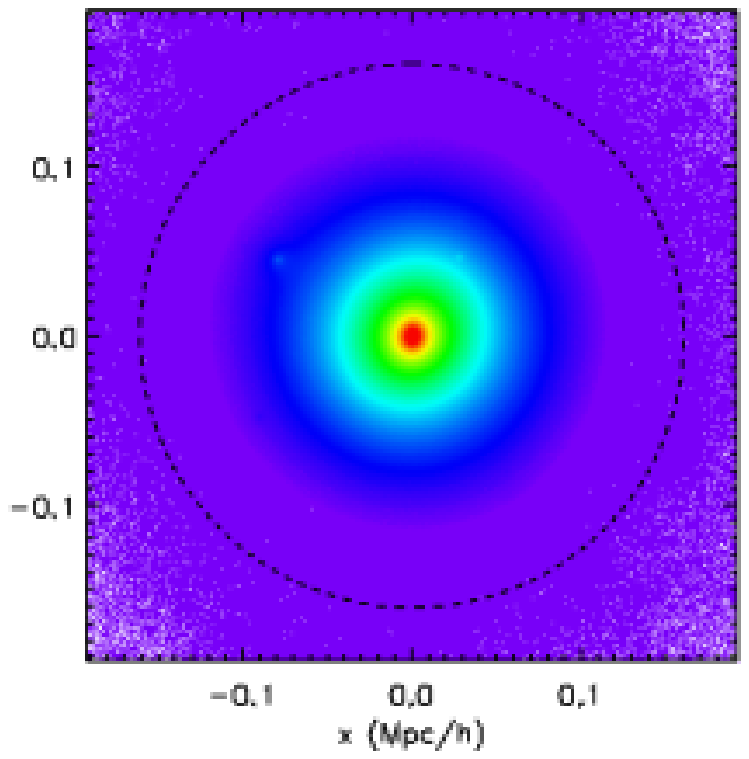}{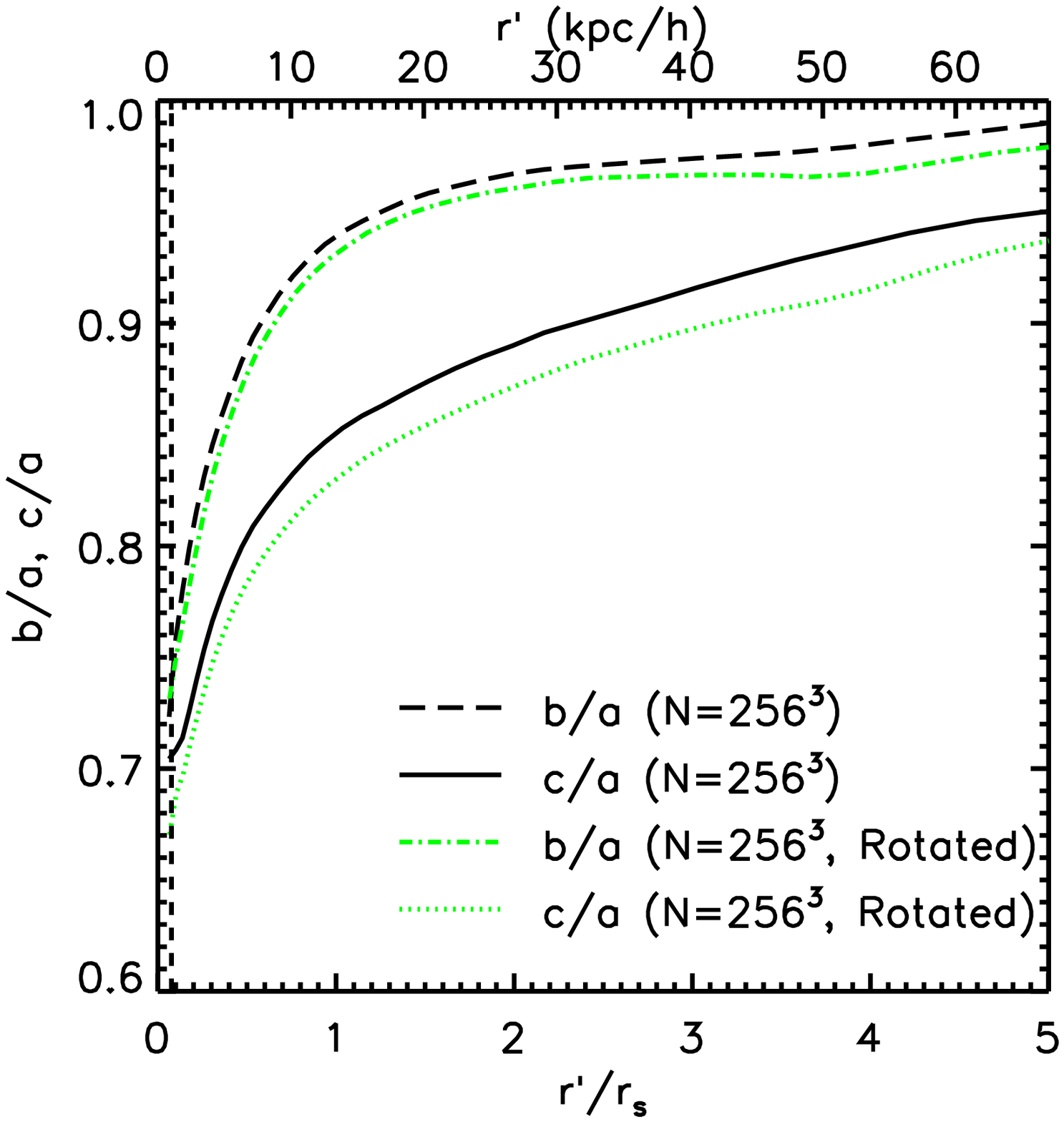}{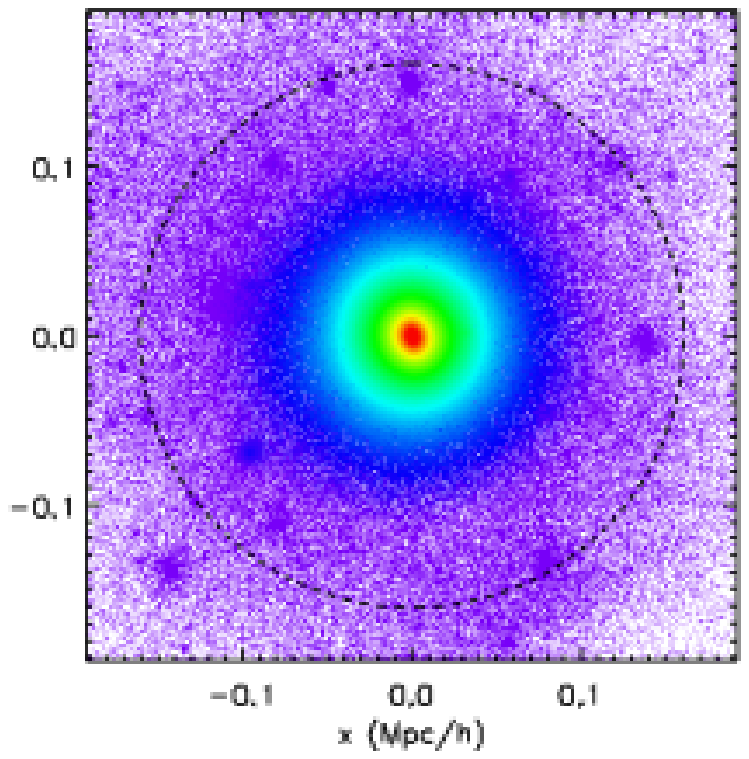}{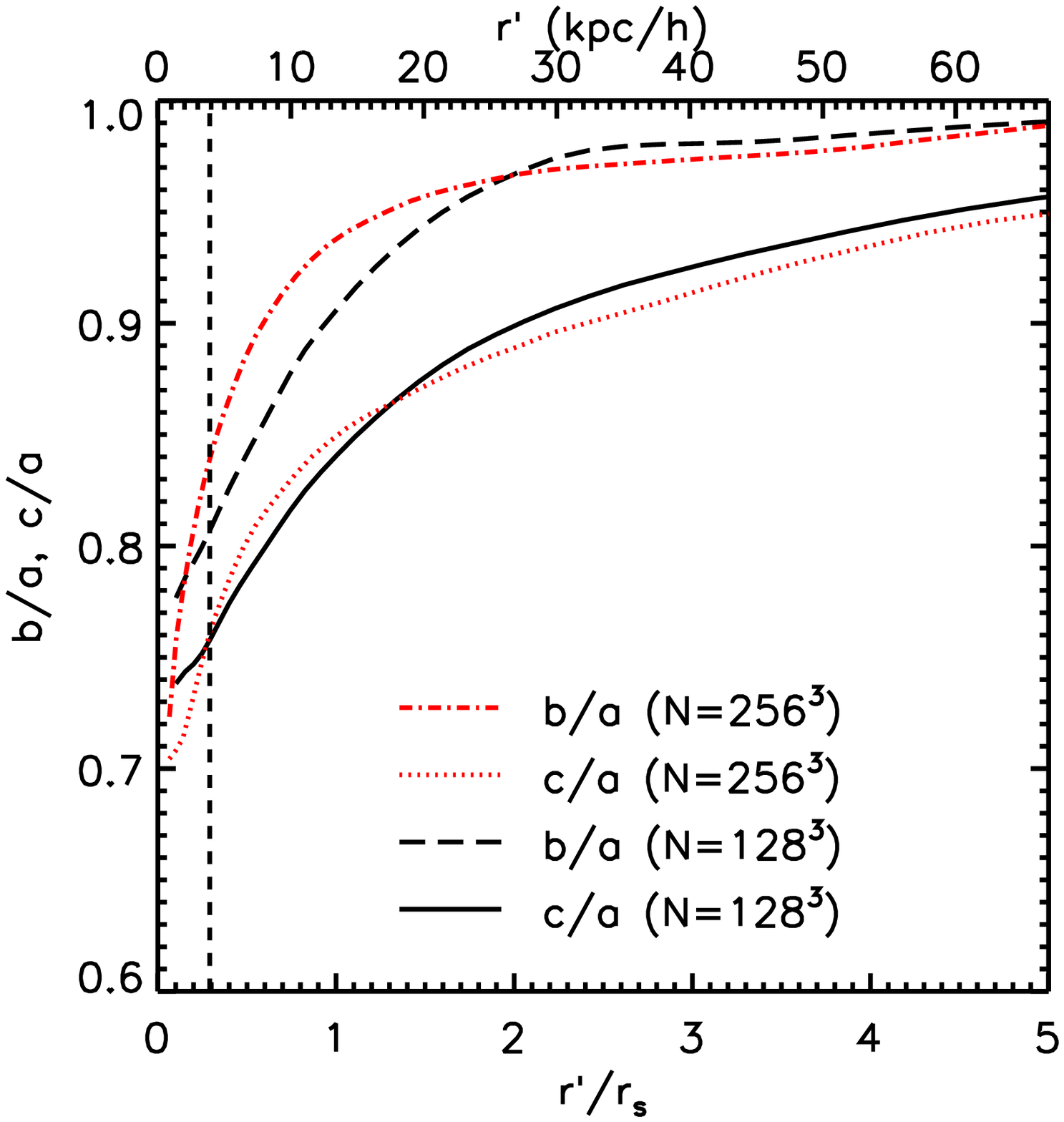}{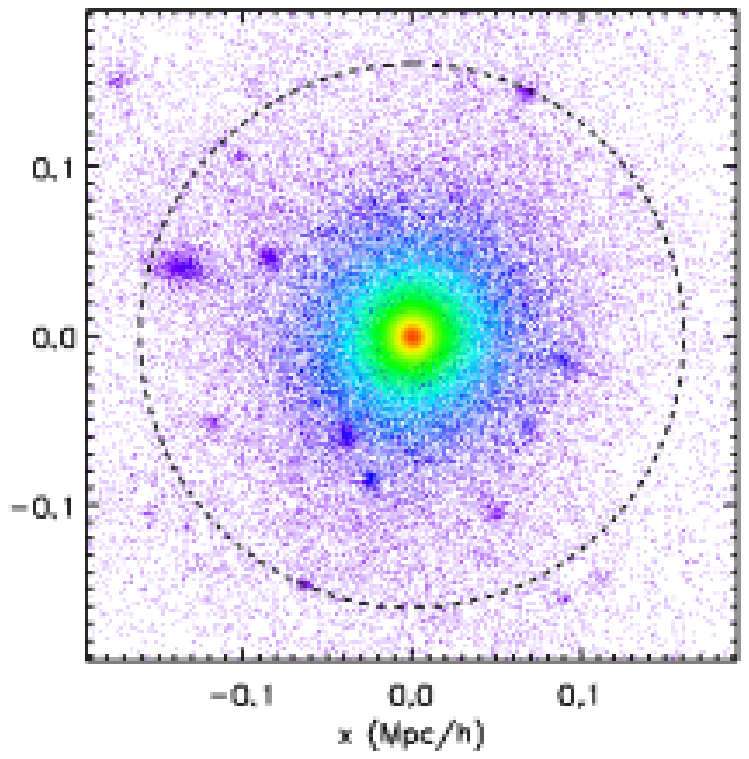}{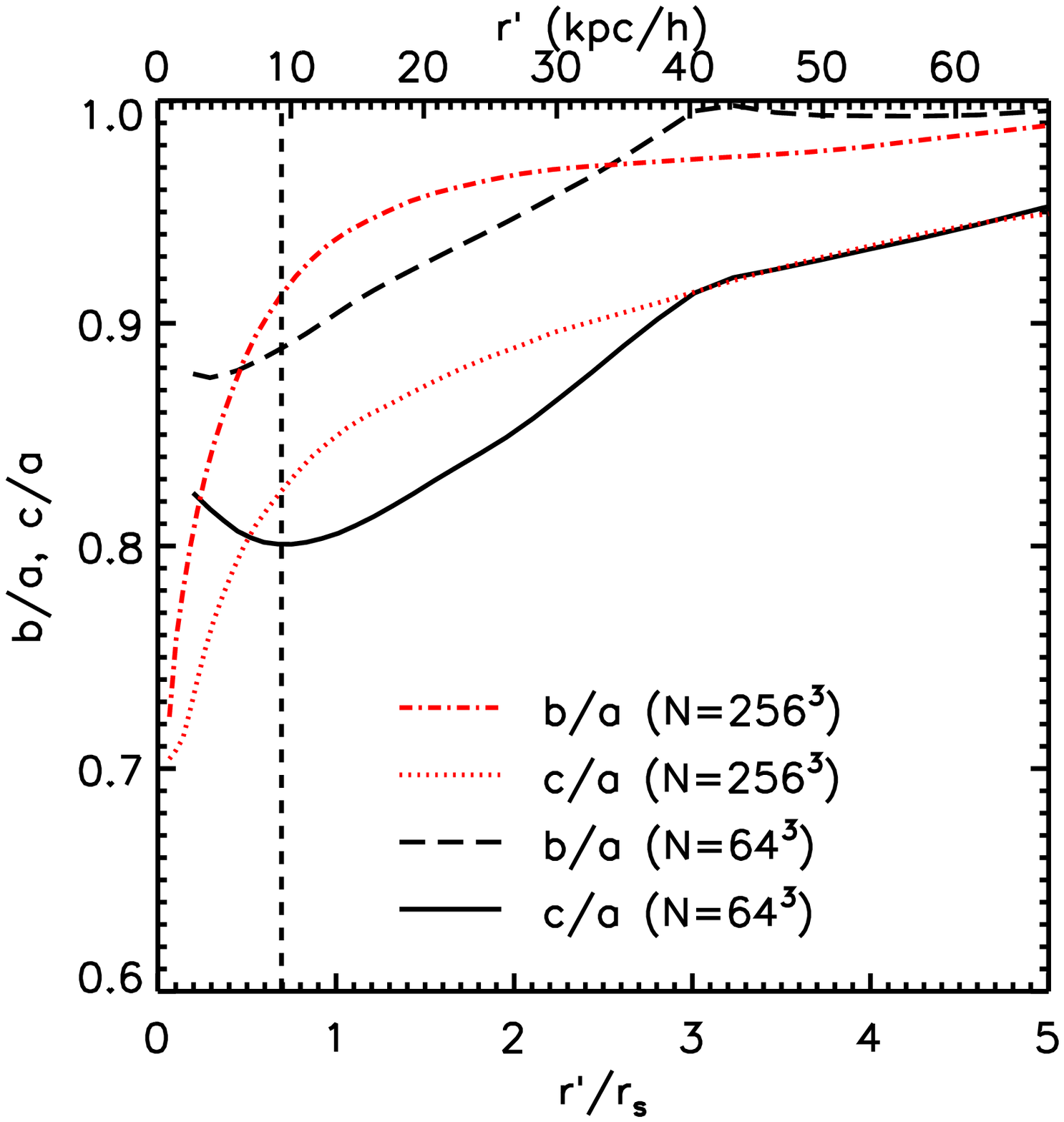}
\caption{ Halo G7 simulated at three different resolutions. The number of
particles varies by a factor of 8 between each realization.  {\it Left panels:}
Halo particles coloured by gravitational potential.  {\it Right panels:} Axial
ratios $b/a$ and $c/a$ measured by fitting ellipsoids to isopotential contours
on three orthogonal planes through the halo as a function of the elliptical
radius $r'=(a^2+b^2+c^2)^{1/2}$.  The vertical dashed line shows the minimum
converged radius in the mass profile, $r_{\rm conv}$, in each simulated halo.
The top right panel shows the axial ratios calculated using two sets of
orthogonal planes rotated by an arbitrary angle.  The axial ratios measured in
both cases are in good agreement and differ by $\lsim~3\%$ at all radii.
\label{fig:convergence}}
\end{figure}

Figure~\ref{fig:convergence} shows the results of this procedure applied to
galaxy halo G7. The right panels show the ratio of the semimajor axes of the
best fitting ellipsoids, $b/a$ and $c/a$, where $c < b < a$, as a function of
the radius $r'=\sqrt{a^2+b^2+c^2}$.  The radius is plotted in units of the NFW
scale radius, $r_s$ (where the logarithmic slope of the density profile is equal
to the isothermal value of $-2$), determined by fitting the spherically-averaged
density profile with the NFW fitting formula.

The top panels show the highest resolution realization, G7/256$^3$,
which has $N_{200}\simeq3.5\times10^6$ particles within $r_{200}$
(shown by a circle in the left panels). Both the minor-to-major
($c/a$) and intermediate-to-major ($b/a$) axial ratios decrease
gradually inwards, with evidence of a sharp decrease inside the scale
radius.  The top right panel also illustrates the uncertainty in the
shape measurements by plotting the axial ratios calculated using two
sets of orthogonal planes rotated by an arbitrary angle.  The axial
ratios measured in both cases are in good agreement and differ by
$\lsim~3\%$ at all radii, confirming the applicability of
approximating the isopotential surfaces as ellipsoids.

The middle and lower panels of Figure~\ref{fig:convergence} show the
results for two lower resolution simulations of the same halo G7.  The
realizations labelled $N=128^3$ and $N=64^3$ have $N_{200} \simeq 3.3
\times 10^5$ and $4.2 \times 10^4$, respectively.  The axial ratios of
G7/$128^3$ are in good agreement with those of the high resolution
$N=256^3$ run, with maximum deviations of $\simeq 4\%$.  In the lowest
resolution run, however, although the general trend in the axial
ratios is reproduced, large deviations are apparent at small
radii. Near the centre, G7/64$^3$ is significantly more spherical than
higher resolution realizations, suggesting that the steep inward
increase in asphericity seen in G7/128$^3$ and G7/256$^3$ is a result
of the inner structure of the halo, which is poorly resolved in the
case of G7/64$^3$.  The two-body relaxation timescale depends on the
number of particles and the gravitational softening length and therefore is
shortest near the centre.  This may result in a more spherical distribution of
particles, which in turn can cause the potential to be more spherical near the
centre of low resolution halos.  We investigate this possibility in
Section~\ref{sec:model}, but for the rest of the analysis we use only
simulations of resolution comparable to the 256$^3$ realization of halo G7.

\section{Halo Shapes and Internal Alignment}
\label{sec:shapes}

\subsection{Radial dependence of isopotential shapes}

Because the monopole dominates in the outer regions of a centrally
concentrated mass distribution, the isopotential surfaces are expected
to become more spherical in the outer regions. The precise radial
variation of the shape depends on the distribution of mass within the
halo and on the radial dependence of its flattening.

\begin{figure}
\plotone{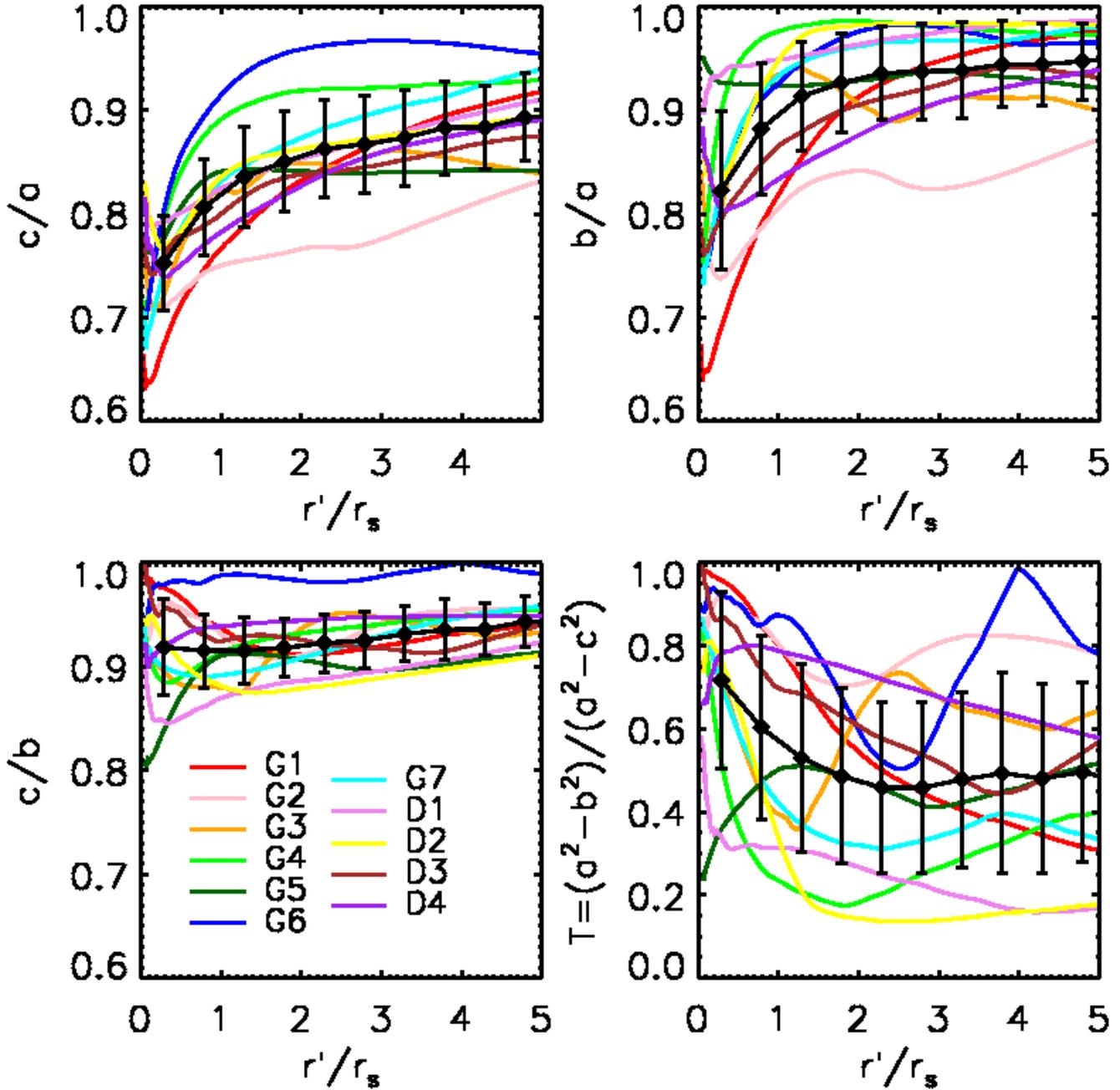}
\caption{Isopotential shapes of all halos as a function of radius,
where $c$, $b$, and $a$, are the lengths of the minor, intermediate
and major axes, respectively, and the radius is given in units of the
NFW scale radius of the density profile, $r_s$.  The triaxiality
parameter $T=(a^2-b^2)/(a^2-c^2)$ is equal to 0 (1) for a perfectly
oblate (prolate) spheroid.  Points with error bars show the mean $\pm
1~\sigma$ of all axial ratio profiles. Halos clearly tend to become
more aspherical towards the centre, a gradual trend that becomes
increasingly steep inside $r_s$. On average, halos are very nearly
prolate at the centre, but become more spherical and less axisymmetric
in the outer regions.
\label{fig:isopot3dplt}}
\end{figure}

Figure~\ref{fig:isopot3dplt} shows the axial ratio profiles for all of
the simulated halos in our sample along with the average profile
(computed after scaling each profile to its scale radius, $r_s$) with
1$\sigma$ error bars to indicate the scatter.  The slow approach to
spherical symmetry in the outer regions is clear in all cases (the
average minor-to major axial ratio at $r = 5~r_s$ is $c/a \simeq
0.9$), as is the sharp inward increase in asphericity inside the scale
radius. Near the centre, $c/a$ and $b/a$ approach similar values,
$0.72$ and $0.78$, respectively, but the system becomes more spherical
and less axisymmetric in the outer regions.
Figure~\ref{fig:isopot3dplt} also shows the triaxiality parameter,
defined as $T=(a^2-b^2)/(a^2-c^2)$, where $T= 0~(1)$ for a perfectly
oblate (prolate) spheroid.  On average, $T > 0.5$ for $r < r_s$
indicating that halos tend toward prolate shapes near the centre.  The
$c/b$ profiles remain relatively flat from the centre outwards,
increasing from a central value of $\sim 0.92$ to $\sim 0.94$ in the
outer parts of the halo.

\begin{figure}
\plotone{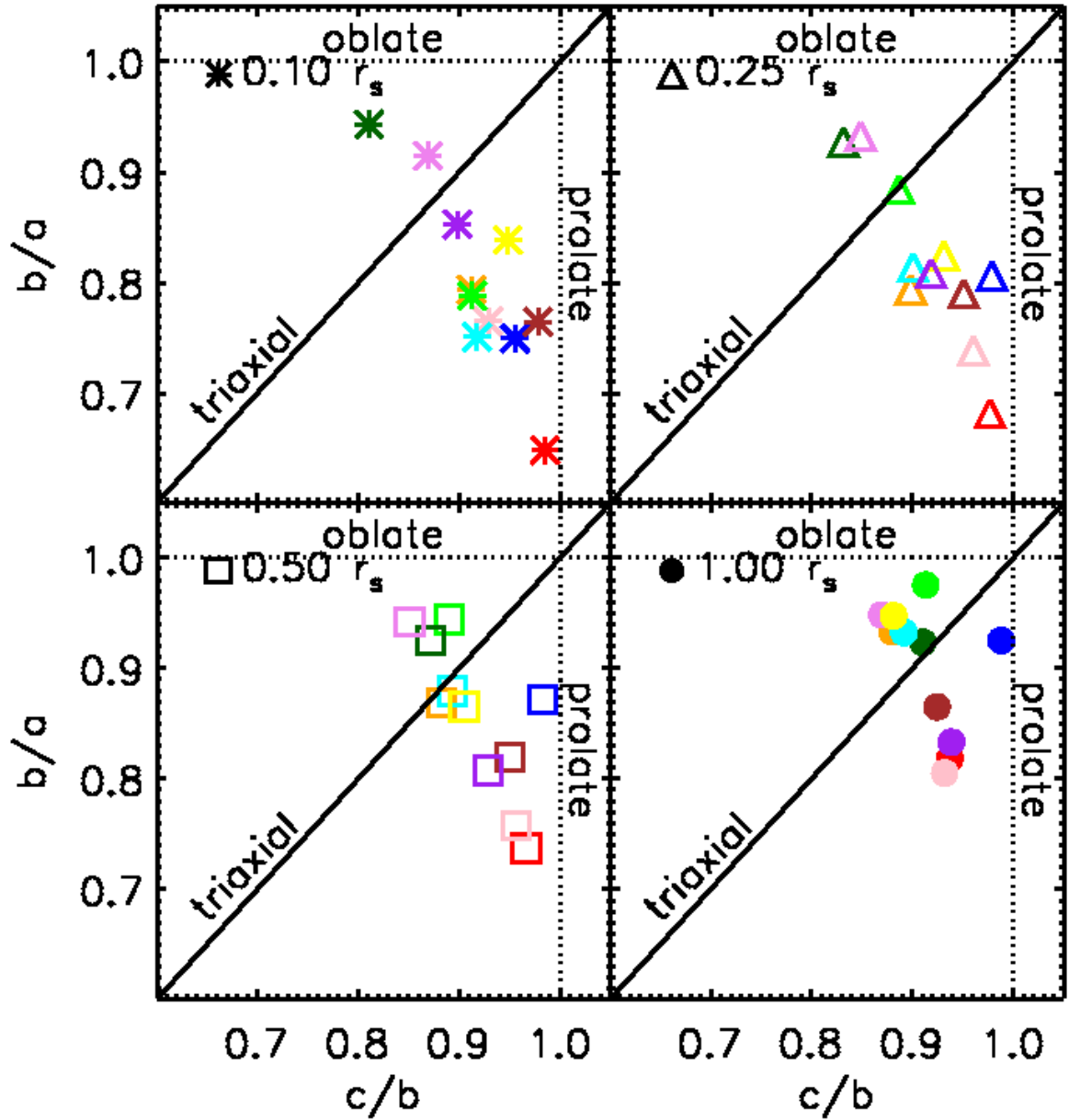}
\caption{Intermediate-to-major ($b/a$) versus minor-to-intermediate
$(c/b)$ axial ratios at four different radii within $r_s$.  Perfectly
oblate (prolate) halos have $b/a=1$ ($c/b=1$) and ``maximally
triaxial'' halos have $b/a=c/b$.  Halos tend to become more
prolate inwards.
\label{fig:isopot3d_ob3}}
\end{figure}

A more detailed view of this inner region is shown in
Figure~\ref{fig:isopot3d_ob3}, where we plot $b/a$ versus $c/b$ at four
different radii, $r=0.1, 0.25, 0.5$, and $1.0~r_s$.  Note that a perfectly
oblate spheroid corresponds to $b/a=1$, whereas a prolate spheroid corresponds
to $c/b=1$.  There is a clear trend for halos to become increasingly prolate
near the centre.  At $r=r_s$, five out of eleven halos are more prolate than
oblate, but at $r=0.1~r_s$, all but two of the halos are more prolate than
oblate.

The significant variation in the shape of the halo potential with radius may
complicate detailed comparisons of the shapes of simulated halos with those
inferred from X-ray and S-Z observations.  In the model of \cite{LEE04}, for
example, this variation is neglected in order to simplify the calculation of the
projected surface brightness profile of a galaxy cluster.  The large range in
axial ratio values at $r \lsim 2~r_s$ suggests that comparing axial ratios
measured at large radii might be a more straightforward test of the distribution
of shapes.  Observing an increase in the ellipticity of X-ray isophotes or S-Z
isodecrement contours towards the centre in individual cluster systems would
also provide strong evidence for the existence of an underlying CDM halo
potential.  

\subsection{Alignments}

\begin{figure}
\plotone{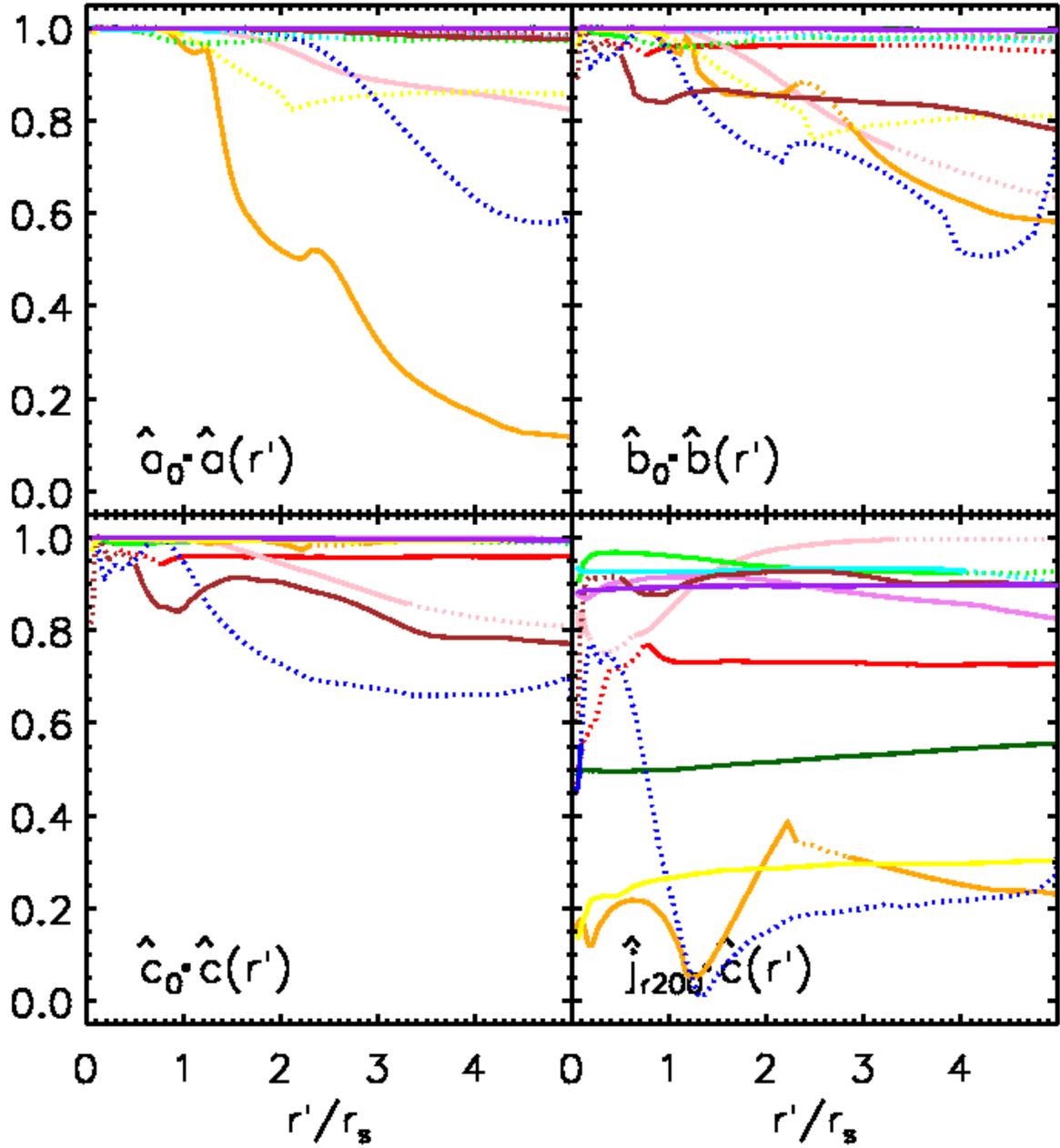}
\caption{ Alignment of principal axes as a function of radius.  Vector
$\hat{a}(r')$ represents the unit vector along the major axis of the
halo, and $\hat{a}_0=\hat{a} (r' \simeq r_{\rm conv})$.  Dotted lines
indicate radii where $c/b > 0.95$ and/or $b/a > 0.95$. For these
nearly axisymmetric systems two of the axes' directions are poorly
determined.  In eight of the eleven halos, the principal axes are well
aligned with radius throughout the main body of the halo. In most
halos $\hat{j}_{r200} \cdot \hat{c}(r) \simeq 0.9$, indicating
alignment of $25^\circ$ or better between the minor axis and the
angular momentum vector of the halo.\label{fig:isopot3dalign}}
\end{figure}

The alignment of the isopotential surfaces as a function of radius is
illustrated in Figure~\ref{fig:isopot3dalign}, which shows the cosine
of the angle between the principal axes of the isopotential at the
``converged'' radius and the corresponding axis at larger radii. Since
orientations between some axes will be poorly determined in cases
where $b \simeq a$ or $c\simeq b$, we plot the results using solid
(dotted) curves in Figure~\ref{fig:isopot3dalign} when the length of
the axis plotted differs by more (less) than $5\%$ from the lengths of
the other axes. Taking this into account, we find that in most cases
where the axes are well determined the isopotential surfaces appear to
be reasonably well aligned as a function of radius, with the possible
exception of halos G2, G3, and D3.

The bottom-right panel of Figure~\ref{fig:isopot3dalign} also shows
the relative alignment between the potential and the angular momentum
of the halo, by plotting the cosine of the angle between the minor
axis and the total angular momentum, computed using all particles
within $r_{200}$.  We find that the minor axis tends to be aligned
with the angular momentum vector in most halos; in six of the eleven
halos, the alignment between the minor axis and the angular momentum
vector is better than $25^\circ$ out to $\simeq 5~r_s$.

The latter result has important implications for the dynamics of disk
galaxies since baryons and dark matter are expected to have acquired
similar (specific) angular momenta, and, therefore, the rotation axis
of a galactic disk would be aligned with the halo's net angular
momentum \citep[see, e.g.,][]{FALL80, VANDENBOSCH02, NAVARRO04A,
ABADI06}. Such a disk would be confined to a plane containing the
major and intermediate axes of the halo.  In a nearly prolate halo the
potential in such plane would deviate significantly from circular
symmetry.

\begin{figure}
\plotfour{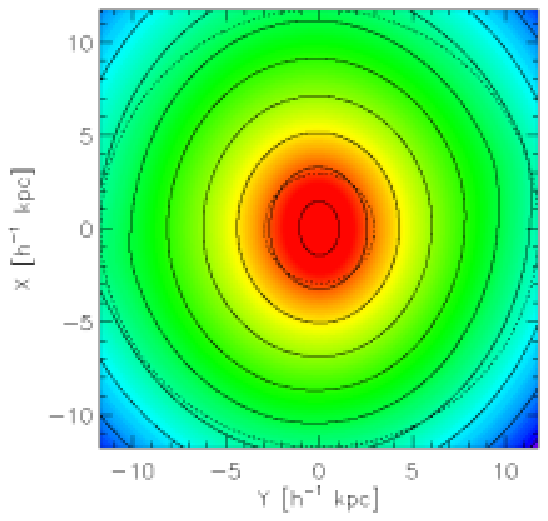}{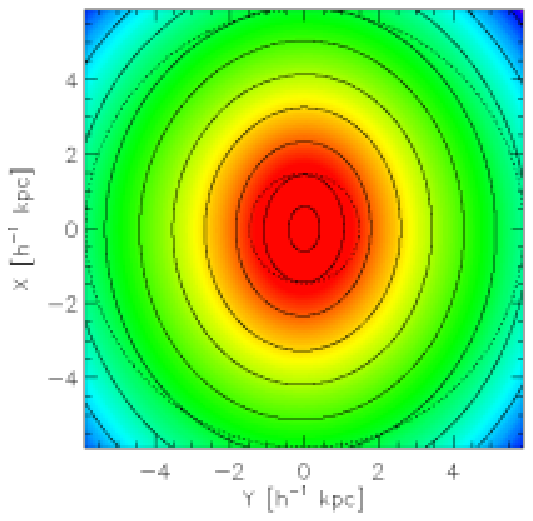}{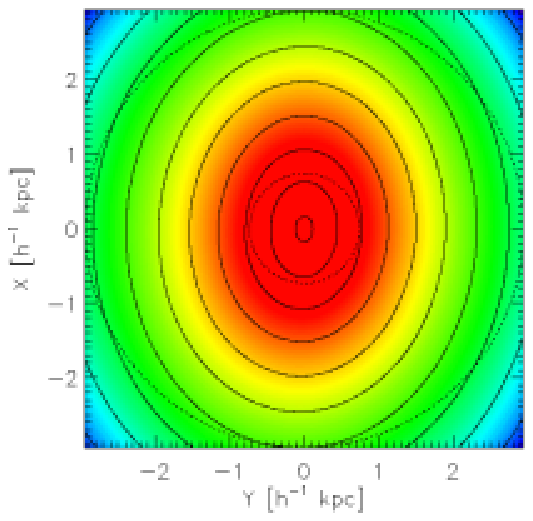}{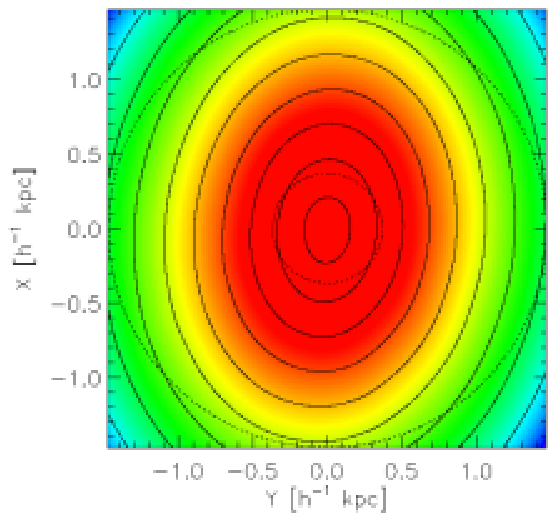}
\caption{Gravitational potential of halo D3 on the plane perpendicular
to its minor axis, corresponding to the plane in which a disk might
form.  The radii of the outer dotted circles are $4,2, 1,$ and
$0.5~r_s$ in the top left, top right, bottom right panels,
respectively, where $r_s = 2.94~\kpch$.  The radius of the inner
circles is $0.25\times$ that of the outer circle.  The ellipticity of
the isopotential contours (solid curves) clearly increases towards the
centre of the halo.
\label{fig:isopotzoom}}
\end{figure}

We show this explicitly in Figure~\ref{fig:isopotzoom} where we plot
the potential of halo D3 on the plane containing its major and
intermediate axes.  The isopotential contours are clearly noncircular
but are well-approximated by ellipses.  The four panels in
Figure~\ref{fig:isopotzoom} show the halo on increasingly small
scales in order to highlight the increase in the ellipticity of the
isopotential contours towards the centre of the halo.  We investigate
the consequences of this for the dynamics of a gaseous disk in
Section~\ref{sec:disk}.

\section{A Simple Model for Halo Triaxiality}
\label{sec:model}

The radial dependence of the shape of the potential is dictated by
both the flattening and the mass profile of the halo.  Here we explore
the effects of each in order to shed light on the steep radial
dependence of the halo asphericity discussed in the previous section.

To this aim, we generate a spherically symmetric NFW halo model with
$N=10^6$ particles out to a maximum radius of $10~r_s$.  We set the
gravitational softening length to $0.05~r_s$. We turn this
into a prolate halo of constant flattening by multiplying the x- and
y-coordinates of all particles by a factor of $0.4$.  We then use the
ellipsoid fitting method described in the previous section to measure
the shapes of the isopotential surfaces as a function of radius.  We
also measure the shape of the mass distribution by diagonalizing the
inertia tensor according to the iterative procedure of \cite{KATZ91}.
This method measures the shape using all particles within an ellipsoid
whose volume is approximately equal to the volume of a sphere of
radius $r$.

\begin{figure}
\plottwo{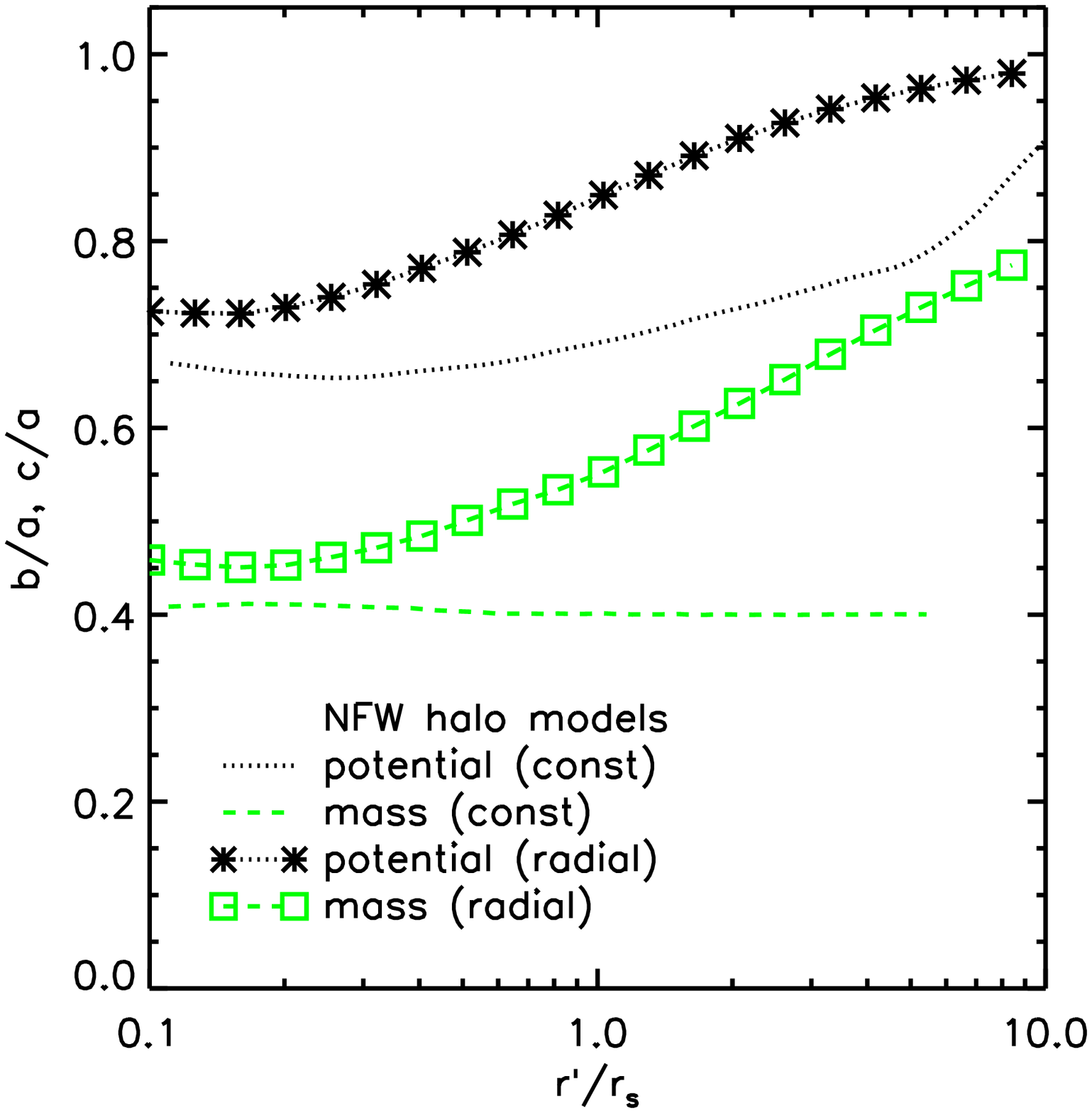}{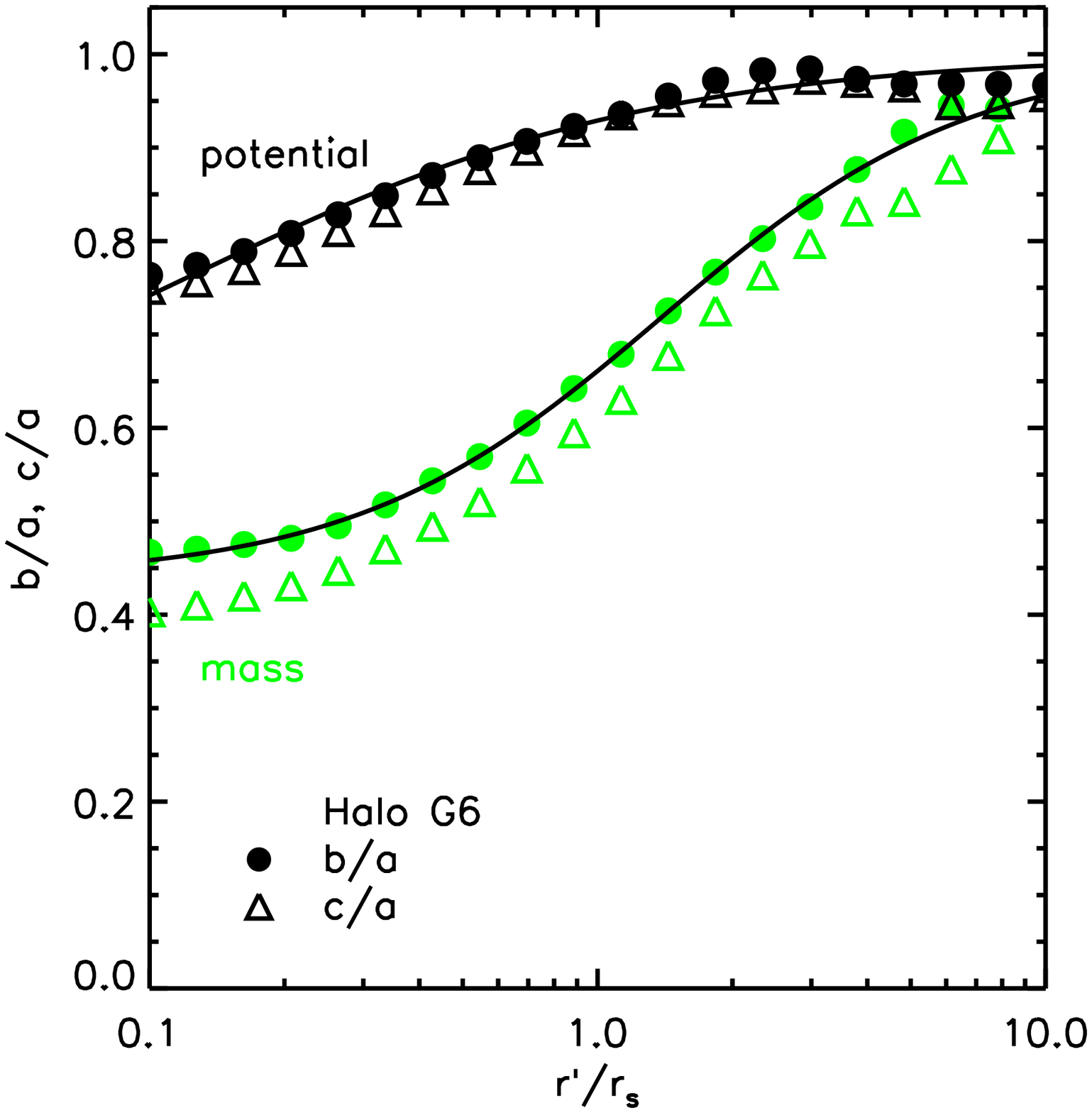}
\caption{ {\it Left panel:} Axial ratios of NFW halo models with
constant (lines without symbols) and radially varying (symbols)
flattening in the mass distribution as a function of radius. {\it
Right panel:} Axial ratios of halo G6 as a function or radius.  Solid
curves show fits to the axis ratio profiles with eq.~(\ref{eq:tanhfit}).
Halo models with constant flattening cannot reproduce the axial ratio
profiles of simulated halos like G6, which require increasing
asphericity towards the centre.
\label{fig:inertisopot}}
\end{figure}

Figure~\ref{fig:inertisopot} compares the shape of the self-similar
prolate NFW halo (left panel) and halo G6 (right panel). The dashed
curve (without symbols) in the left panel shows the axial ratio
measured using the inertia tensor; as expected, a constant value of
$\approx 0.4$ is recovered for the prolate NFW model. The potential
axial ratio (dotted curve without symbols) is not constant, but the
radial dependence is much gentler than seen in halo G6. Within $r_s$
the shape of the potential of the prolate NFW model remains more or
less constant at $\sim 0.65$; and becomes gradually more spherical
outside $r_s$. This differs from the radial behaviour of the axial
ratios in halo G6 (right panel) where both the potential and inertia
shapes within $r_s$ increase from the centre outwards.  These
results indicate that the sharp increase in the asphericity of the
halo potential is inconsistent with a constant flattening in the mass
distribution and reflects the presence of a radially varying
distortion in the mass distribution as well.

A convenient approximation of the shape profiles is provided by the
following formula:
\begin{equation}
\log \left({b\over a} \ {\rm or}\ {c\over a}\right) = \alpha \left[ \tanh \left(
  \gamma \log \frac{r}{r_{\alpha}} \right) -1 \right].
\label{eq:tanhfit}
\end{equation}
Here $\alpha$ parameterizes the central value of the ratio, such that
$(b/a)_0$ or $(c/a)_0$ is given by $10^{-2~\alpha}$, $r_{\alpha}$
indicates the characteristic radius at which the axial ratio rises a
significant amount from its central value, and $\gamma$ regulates the
sharpness of the transition.  This fitting formula gives a reasonably
accurate description of the axial ratio profiles of both the mass
distribution and the isopotential surfaces in the simulations, as
shown by the fits to the G6 shape profiles in
Figure~\ref{fig:inertisopot} (lines through symbols in its right
panel).  The best fit parameter values for fits to the mass and
isopotential shapes are given in Table~\ref{tab:halos}.

An NFW halo model with a radially varying flattening is a much better
approximation to the radial behaviour of the shapes seen in the
simulations. This is shown in the left panel of
Figure~\ref{fig:inertisopot}, where the same (initially spherical) NFW
model is modified by multiplying the x- and y- coordinates of
particles at radius $r$ by eq.~(\ref{eq:tanhfit}) evaluated with
($\alpha, r_{\alpha}, \gamma$) = $(0.2, 1.0~r_s, 1.3)$. This choice
%($\alpha, r_{\alpha}, \gamma$) = $(0.18, 1.0~r_s, 1.4)$. This choice
is motivated by the best fit to the $b/a$ profile of the mass
distribution of halo G6.

The axial ratio profiles of the resulting halo model are shown by the
symbols in the left panel of Figure~\ref{fig:inertisopot}.  The
distortion in the shape of the isopotential surfaces of this halo is
clearly reminiscent of that of halo G6. The agreement is not exact,
however, because the inertia tensor measures the shape using all
particles within a given volume and therefore does not correspond to a
transformation applied to the coordinates of particles as a function
of radius. However, the model does demonstrate that a distortion in
the mass distribution which increases towards the centre of the halo
can reproduce the radial variation observed in the isopotential shapes
of halos like G6. 

We also find that in all cases, the potential axial ratio profile becomes flat,
or increases slightly, toward the centre of the halo.  In G6 and the radially
varying NFW model, this coincides with a similar trend in the mass axial ratio
profile.  However, in the NFW model with a constant flatenning, the shape of the
potential becomes more spherical near the centre, but the shape of the mass
distribution remains constant at all radii.  This suggests that the finite
gravitational softening length may be responsible for making the potential more
spherical near the centre, although two-body relaxataion effects may also play a
role near the centres of simulated halos like G6 and the low resolution versions
of G7 shown in Figure~\ref{fig:convergence}. 

We note that the ratio of the potential axial ratio to the mass axial ratio
varies as a function of radius for both the constant and radially varying
flatenning models.  In terms of the eccentricity parameter used in the models of
\cite{LEE03} and \cite{LEE04}, $e=\sqrt{1-(b/a)^2)}$, the ratio of potential
eccentricity to the mass eccentricity varies from $\sim0.8$ to $\sim0.6$, and
$\sim0.8$ to $\sim0.3$, in the constant and radially varying cases,
respectively.  In comparison, the eccentricity ratio of the \cite{LEE03} model
varies from $0.77$ to $0.4$, however, \cite{LEE04} approximate this with a
constant value of $0.49$ in order to calculate the projected surface brightness
distribution of their model.  Since this ratio varies over roughly the same
range for the model with a radially varying flatenning, we conclude that the
assumption of a constant ratio is no worse in this case than it is for models
with constant flatenning.

\begin{figure}
\plotone{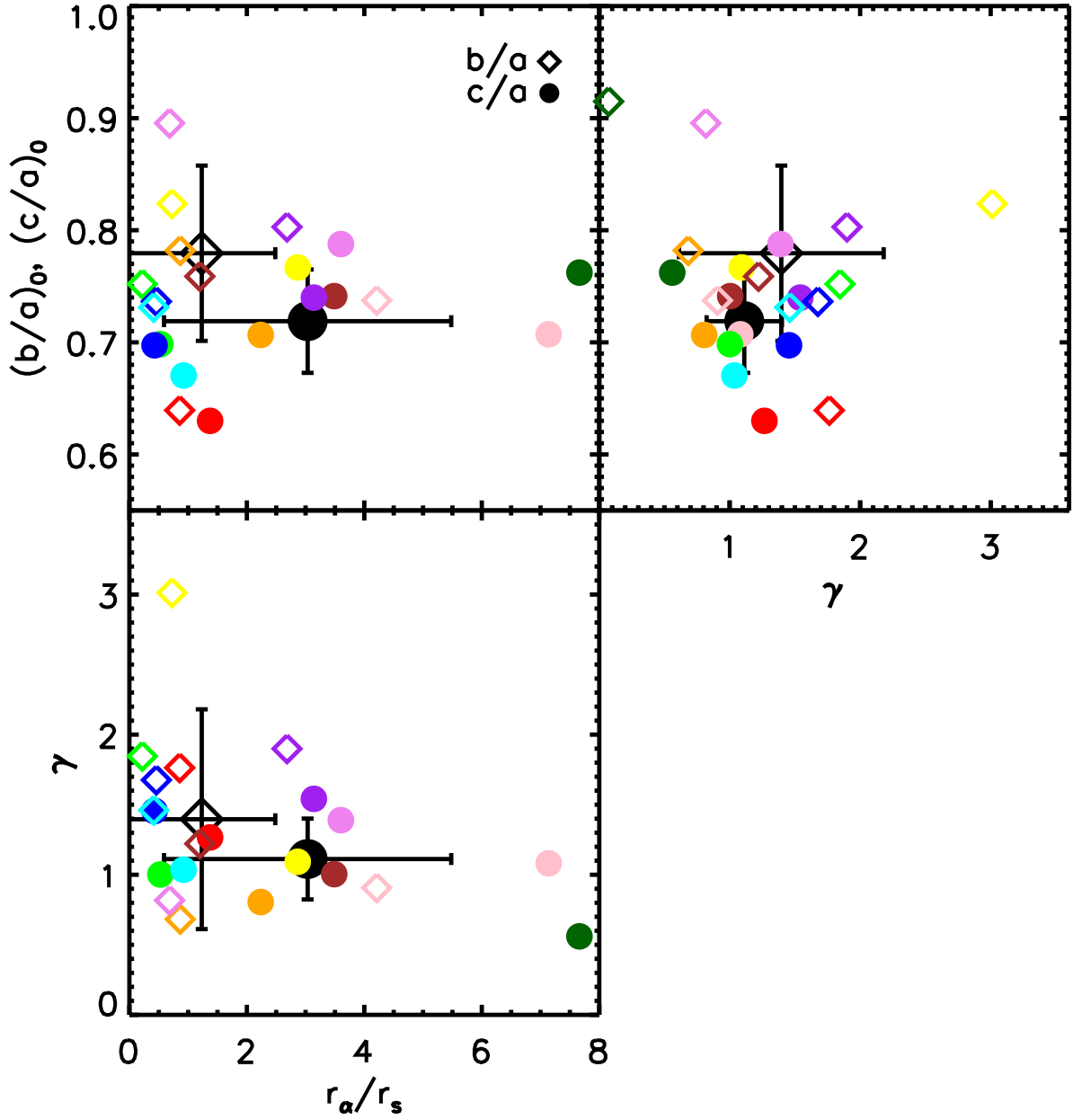}
\caption{ Values of central axial ratios, $(b/a)_0$ and $(c/a)_0$, and
parameters $r_{\alpha}$ and $\gamma$ obtained by fitting
eq.~(\ref{eq:tanhfit}) to the $b/a$ and $c/a$ axial ratio profiles
plotted in Figure~\ref{fig:isopot3dplt}.  Large symbols with error
bars show the average and standard deviation for all halos. The
average transition scale for the $c/a$ profiles is about twice that of
the $b/a$ profiles, indicating that the former tend to increase more
gradually than the latter profiles
\label{fig:cabafitpars}}
\end{figure}

The best fit parameters using eq.~(\ref{eq:tanhfit}) to approximate the
potential axial ratio profiles of all our simulated halos are listed
in in Table~\ref{tab:halos} and shown in Figure~\ref{fig:cabafitpars}.
Although there is substantial halo-to-halo scatter, several trends are
apparent.  The large symbols with ``error bars'' in
Figure~\ref{fig:cabafitpars} show the average value of each parameter
and the 1-sigma scatter. The central axial ratios scatter about
$(b/a)_0=0.78\pm0.08$ and $(c/a)_0=0.72\pm0.04$.  The average
transition radius, expressed in units of the scale radius,
$r_{\alpha}/r_s$, is $1.2 \pm 1.2$ for the $b/a$ profiles (excluding
halo G5 which has an almost flat $b/a$ profile, and therefore an
extremely large and poorly defined value of $r_{\alpha}$) and $3.0 \pm
2.4$ for the $c/a$ profiles.  The average value of $\gamma$ is $1.4
\pm 0.8$ for the $b/a$ profiles and $1.1 \pm 0.3$ for the $c/a$
profiles.  On average, $c/a$ profiles tend to increase more gradually
than $b/a$ profiles, and profiles become more spherical and less
axisymmetric at larger radii.

Our results may be used to construct a model for the gravitational
potential of a triaxial CDM halo by using eq.~(\ref{eq:tanhfit}) to
modify the potential of a spherical NFW model.
\begin{eqnarray}
\label{eq:phitriax1}
\Phi(x,y,z) &= &\Phi_{\rm NFW} (r^{t}) \\
\label{eq:phitriax2}
r^{t} &= & r\sqrt{ \left(\frac{x}{a(r)}\right)^2 + \left(\frac{y}{b(r)}\right)^2 + \left(\frac{z}{c(r)}\right)^2}\\
\label{eq:phitriax3}
\Phi_{\rm NFW}(r) &= &-\frac{G M_{200}}{r_s f(c_{200})} \frac{\ln (1 +
r/r_s)}{r/r_s},
\end{eqnarray}
where the concentration parameter $c_{200}=r_{200}/r_s$ and
$f(u)=\ln(1+u)-u/(1+u)$.  Since eq.~(\ref{eq:tanhfit}) describes only
the ratios between the lengths of the principal axes, the
normalisation is set by assuming that at each radius $r$, the volume
of the ellipsoid with semi-axis lengths $a(r), b(r)$ and $c(r)$ is
equal to $4\pi r^3/3$, i.e., $a(r)b(r)c(r) = {r^3}$, which implies
\begin{equation}
a(r)=r \ (b/a)^{-1/3}(c/a)^{-1/3},
\end{equation}
where $b/a$ and $c/a$ are functions of $r$ given by
eq~\ref{eq:tanhfit}.  This ensures that the spherically-averaged
potential of the triaxial model will be similar to that of a
spherically symmetric NFW model with the same mass and scale radius.

\begin{figure}
\plottwo{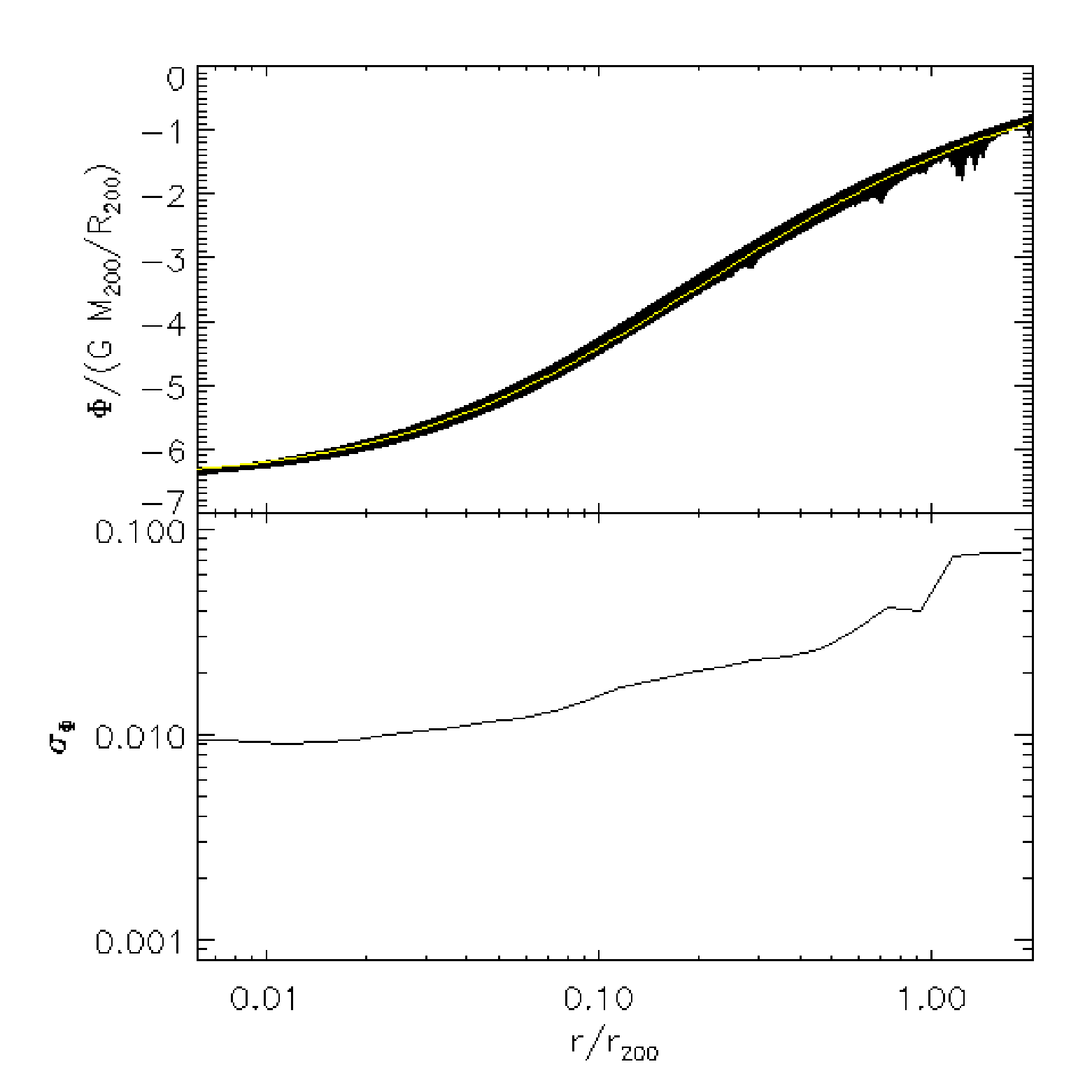}{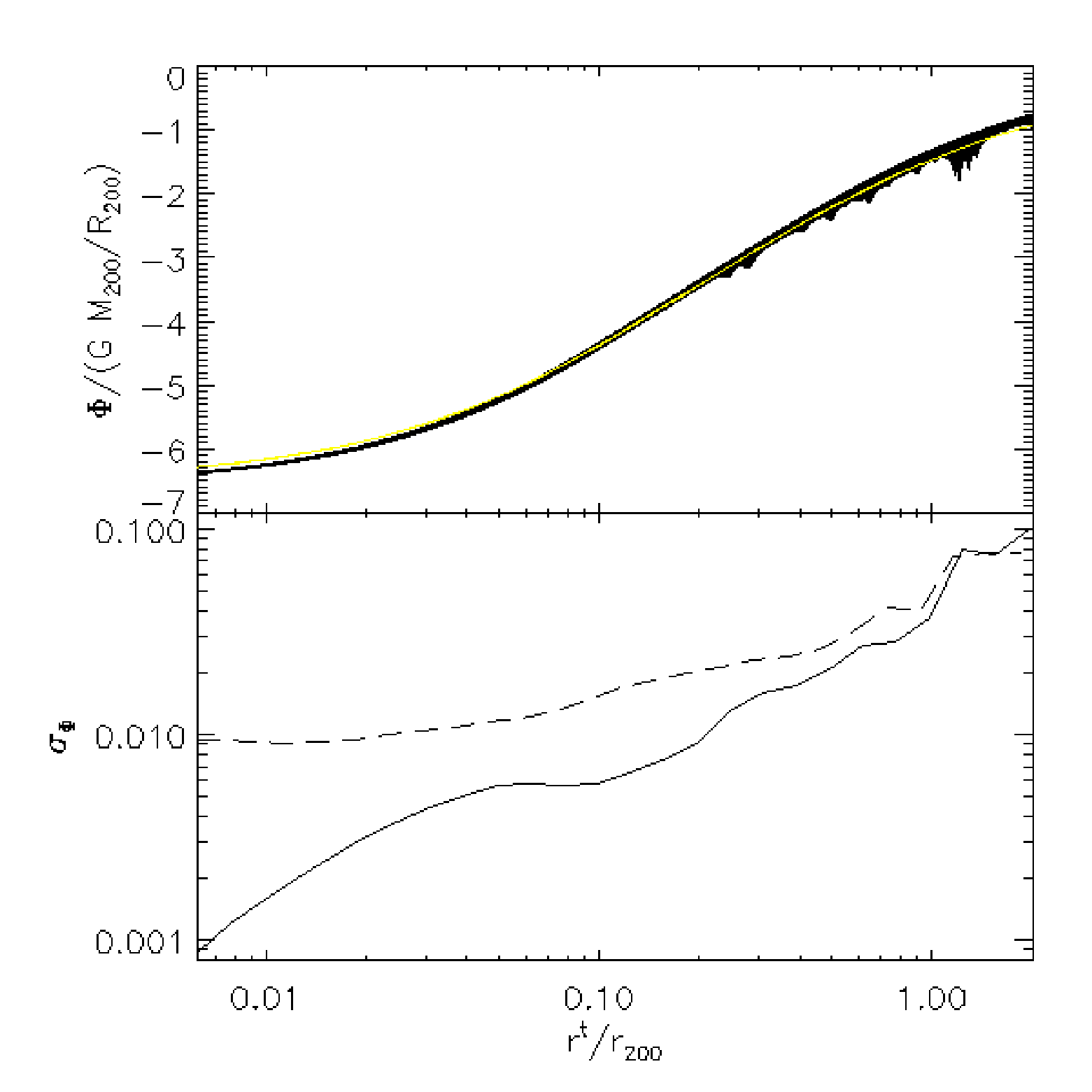}
\caption{{\it Left panels:} Potential versus radius of particles in
halo G5.  The solid line shows the profile of a best fit NFW
potential.  The lower panel shows the residuals from the fit,
$\sigma_\Phi$ as a function of radius.  {\it Right panel:} Same as
left but plotted versus the reduced radius $r^t$ of the best fitting
ellipsoid (see eq.~\ref{eq:phitriax2}). The dashed line in the lower
panel indicates the same residuals, $\sigma_\Phi$, as in the left
panel.  The residuals are significantly reduced at small radii where
substructure does not dominate the scatter.\label{fig:pospotplt_un}}
\end{figure}

We show that this model provides a good description of the potential
of simulated halos in Figure~\ref{fig:pospotplt_un} where we compare
the gravitational potential of particles in halo G5 plotted against
their radius, $r$, and against the reduced radius, $r^{t}$, given by
eq.~(\ref{eq:phitriax2}).  The reduced radius is calculated by rotating
the halo so that its principal axes are aligned with the x-, y-, and
z-coordinate axes. The bottom panels of this figure show that the
dispersion in the potential profile is significantly reduced when
expressed in terms of the reduced radius; in the inner regions, $r
\lsim 2~r_s$, the dispersion is reduced by a factor of two.  At larger
radii, the presence of substructure dominates the scatter in the potential
and the triaxial modelling has less impact.

We conclude that taking into account halo triaxiality provides a substantial
improvement over spherically symmetric CDM halo models.  The fitting formula
given by eq.~(\ref{eq:tanhfit}) along with parameter values in the ranges we
find for simulated halos, listed in Table~\ref{tab:halos}, provides a simple
analytic model for a realistic, triaxial CDM halo potential.  This model may be
of practical use for N-body simulations of tidal streams like those of
\cite{IBATA01} and \cite{HELMI04}, and in calculations of the X-ray and S-Z
surface brightness distributions of triaxial galaxy clusters as in \cite{LEE04}.

%This model may be useful for
%N-body simulations of tidal streams like those of \cite{IBATA01} and
%\cite{HELMI04}, which generally assume a potential model with constant
%axis ratios.

% {Eric: we should perhaps provide at least the average fit parameters
% for the {\it mass} flattening, since that is what is people may want
% to use to create simple models by reshaping spherical ones.??}

\section{Disk Kinematics in Triaxial CDM Halos}
\label{sec:disk}

We explore now the kinematics of disks in CDM halos with radially
varying triaxiality. We assume that the disk lies in one of the
principal planes of the potential, and therefore the problem is
reduced to two dimensions. Our approach follows closely the formalism
presented in HN, where analytic closed loop orbit solutions are found
to describe the orbits in a thin, filled, gaseous disk.  These
solutions use the epicyclic approximation to find closed orbits in a
potential of the form:
\begin{equation}
\Phi(R,\phi) = \Phi_0(R) + \Phi_m(R) \cos(m \phi_0)
\label{eq:potpert} 
\end{equation}
where $\Phi_0(R)$ is the unperturbed (circularly-symmetric) potential,
$\Phi_m(R)$ is a stationary perturbation to that potential, and $\phi$
is the azimuthal angle. For $m=2$ and small perturbations, the
perturbation is periodic in $180^\circ$ and the isopotential contours
are roughly elliptical in shape.  The closed loop orbits solutions are
also roughly elliptical, but are oriented with their major axes
perpendicular to those of the isopotential contours.  

The tangential velocity along the orbit oscillates about the circular
velocity of the unperturbed potential, $V_c(R) = ({R \, {\rm d}
\Phi_0/{\rm d} R})^{1/2}$, and the deviations from $V_c$ will be maximal at
pericentre and apocentre of the orbit. Long-slits aligned with these
directions would yield rotation curves whose shape will systematically
deviate from the true circular velocity profile.

HN showed that even small perturbations to the potential (i.e., $\Phi_m
\ll \Phi_0$) may result in substantial deviations to the orbital shapes
and thus to the tangential velocities; rotation curves that differ
from the halo's $V_c$ profile may thus be matched by choosing a
specially tailored function $\Phi_m(R)$. In particular, HN compute the
perturbation needed to match a pseudo-isothermal velocity profile
(corresponding to a isothermal sphere with a constant density core) in
a cuspy NFW profile.

\begin{figure}
%\plottwo{figs/epspot.eps}{figs/lgvlosr.ps}
%\plotthreesquare{figs/fig10a.eps}{figs/fig10b.eps}{figs/fig10c.eps}
\plotfour{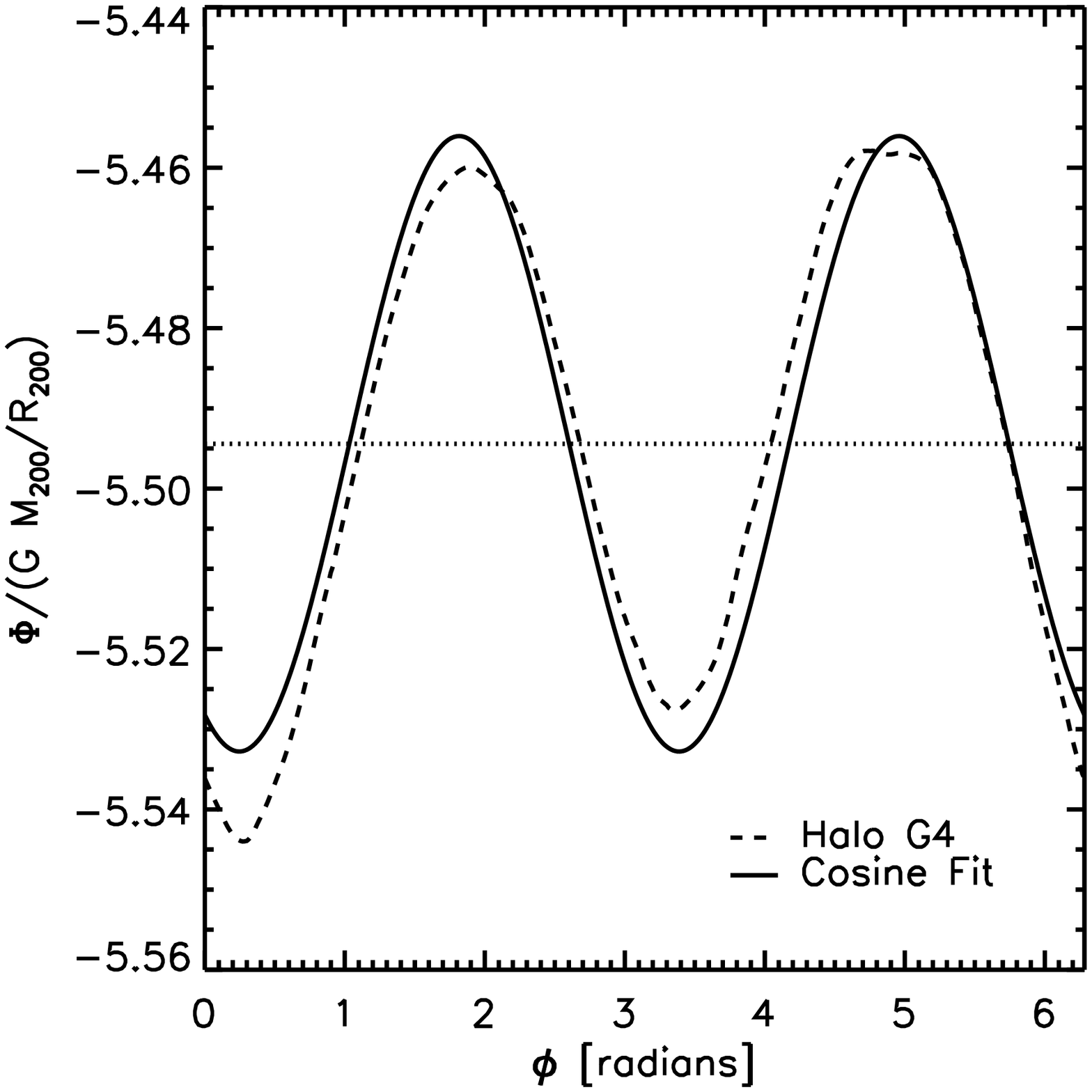}{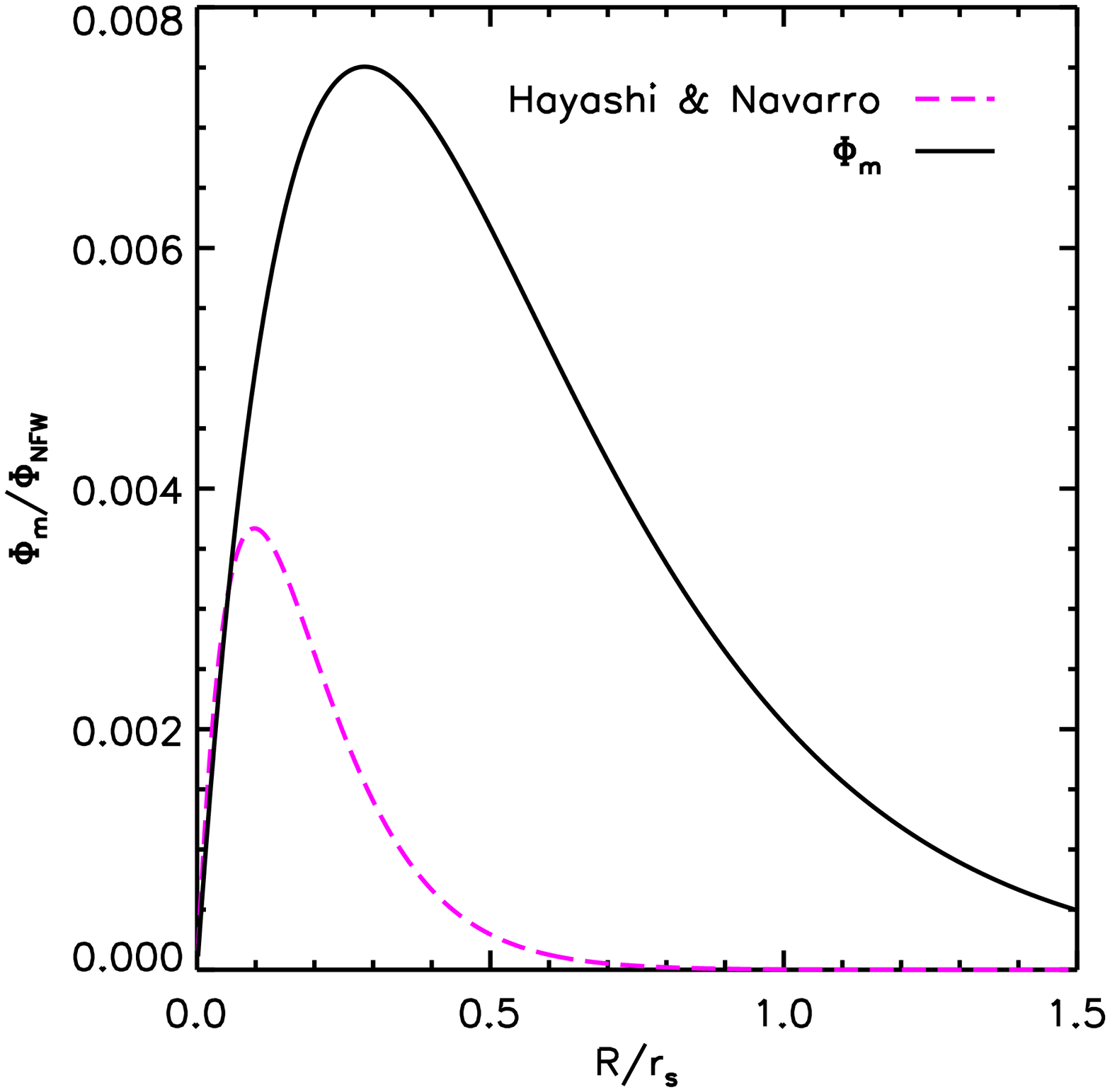}{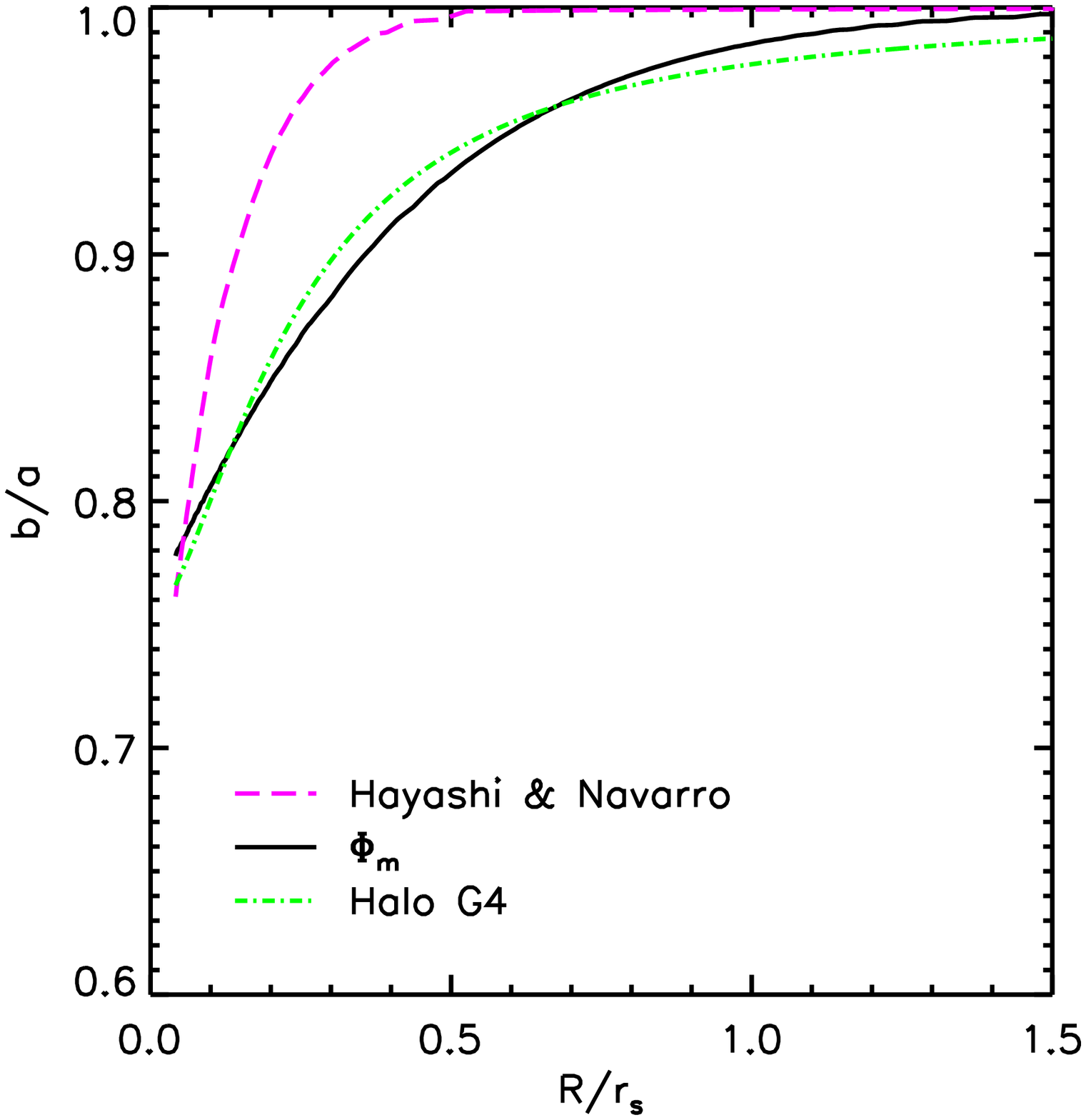}{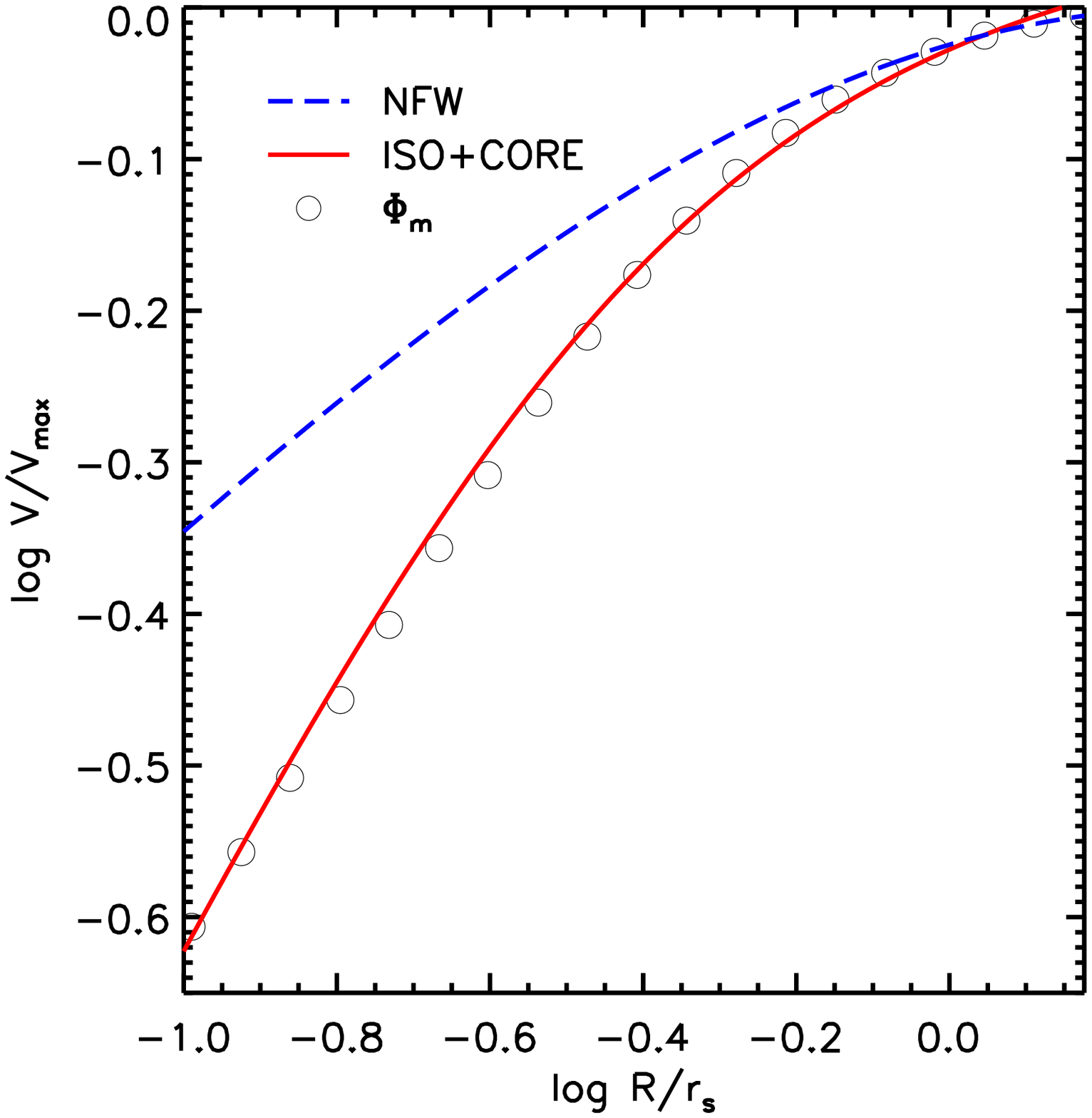}
\caption{ {\it Upper left panel:} The potential plotted versus azimuthal angle
for halo G4 on a ring of radius $0.5~r_s$.  The potential is well fit by a
sinusoidal function whose mean value and amplitude reflect the magnitude of the
perturbation relative to a spherically-symmetric potential.  {\it Upper
right panel:} The magnitude of the ``perturbing'' potential required to yield
``core-like'' long-slit rotation curves in NFW halos, as presented by Hayashi \&
Navarro (2006) (dashed curve).  The dot-dashed curve corresponds to the
perturbation calculated for halo G4, as measured in the plane that
contains the major and intermediate axes.  The solid curve represents a fit to
this perturbation using eq.~(\ref{eq:fisofit}).  {\it Lower left panel:}
Axial ratios of isopotential contours as a function of radius. The dashed,
dot-dashed, and solid curves correspond to the HN solution, halo G4, and the fit
to halo G4, respectively. {\it Lower right panel:} Rotation curve of a disk in
the perturbed potential given by the fit to halo G4, produced when a slit
samples velocities near the major axis of the disk (open circles).  The dashed
line shows the circular velocity profile corresponding to the unperturbed,
spherically symmetric NFW halo.  The solid line shows the best fitting
pseudo-isothermal profile with a constant density core.
\label{fig:rotcurve}}
\end{figure}

This is shown by the dashed curve in the upper right panel of
Figure~\ref{fig:rotcurve}.  The perturbation has a characteristic
shape: it increases inwards to a well defined peak that occurs well
inside the scale radius of the NFW halo, before decreasing nearer the
origin. This behaviour can be approximated by the following fitting
formula:
\begin{equation}
\left|\frac{\Phi_m}{\Phi_{\rm NFW}}\right|	 = a_m~\left({{R}\over {r_m}}\right) \exp\left(- \frac{R}{r_m}\right),
\label{eq:fisofit} 
\end{equation}
where $r_m\sim 0.098 \, r_s$ and $a_m \sim 9.8\times 10^{-3}$. Note
that $r_m \ll r_s$, as required to affect the rising part of the NFW
rotation curve (which peaks at $\sim 2\, r_s$), and that the overall
perturbation is relatively minor (everywhere less than $0.4\%$).

Is this perturbation consistent with the radially-varying triaxial
structure of CDM halos described in the previous section?  We calculate $\Phi_m$
for a simulated halo with the following procedure.  We compute the potential on
concentric rings on the plane containing the intermediate and major axes (the
most likely one to contain the disk given the alignment between angular momentum
and minor axis).  The top left panel of Figure~\ref{fig:rotcurve} shows the
potential plotted versus azimuthal angle for halo G4 on a ring of radius
$0.5~r_s$.  The potential is periodic in $\pi$ radius, and is reasonably well
fit by a sinusoid with three free parameters which describe its phase, amplitude
and mean value, i.e., $A \cos(2\phi+\phi_0)+ B$.  At each radius, we fit such a
sinusoid to the potential and use the best fit parameters to calculate the
magnitude of the perturbation, $|\Phi_m/\Phi_{\rm NFW}| = |A/B|$.  The result is
shown for halo G4 as the dot-dashed in the top right panel of
Figure~\ref{fig:rotcurve}.  

The magnitude of the perturbation peaks at $r \simeq 0.2~r_s$ and is about $2.5$
times as large as the minimum needed to reconcile a ``core-like'' rotation curve
with a cuspy NFW profile.  Unlike the solution presented in HN, however, the
magnitude of the perturbation for halo G4 does not decrease to zero at large
radii and in this case the closed orbit solutions based on the epicyclic
approximation break down.  We approximate the perturbation of halo G4 by fitting
it over the range $R<0.5~r_s$ with eq.~(\ref{eq:fisofit}).  This fit is shown as
the solid line labelled $\Phi_{\rm fit}$ in the top right panel of
Figure~\ref{fig:rotcurve}, and the fit parameter value are $a_m=0.027$ and
$r_m=0.21~r_s$.

The bottom left panel of Figure~\ref{fig:rotcurve} shows the $b/a$ axial ratio
profiles corresponding to halo G4 and the perturbed NFW potential models. The
axial ratio profile of the $\Phi_{\rm fit}$ perturbation is quite similar to
that of halo G4 and is well fit by eq.~(\ref{eq:tanhfit}) with parameter values
($\alpha, r_{\alpha}, \gamma$) = $(0.079, 0.24~r_s, 2.68)$, compared to $(0.062,
0.22~r_s, 1.85)$ for halo G4.  

The lower right panel of Figure~\ref{fig:rotcurve} shows as open circles the
rotation curve that would be obtained by a long slit sampling tangential
velocities along the major axis of the disk ($\phi_a=0^\circ$ in the notation of
HN) in the perturbed potential given by the fit to halo G4. The slope of the
inner rotation curve is significantly modified from that of the initial NFW
profile, and is well fit by a pseudo-isothermal velocity profile (solid line),
given by $V_{\rm iso}^2(r) = V_\infty^2 [1 - ({r_{\rm core}}/{r})
\tan^{-1}({r}/{r_{\rm core}})]$, where $V_\infty$ is the asymptotic velocity and
$r_{\rm core}$ is the radius of the constant density core in this model.  The
best fitting value of the core radius in this case is $r_{\rm core}=0.38~r_s$,
comparable to the characteristic radius of the perturbation, $r_m\approx
0.45~r_{\alpha} \approx 0.24~r_s$.

%% {Eric: it might be worth looking into this a bit more....are core
%% radii inferred from observations within the ball park of what would be
%% ``predicted'' by our average triaxial profiles for halos of average
%% concentrations? 

%% We should also give the average proportionality factor
%% between $r_m$ and $r_{\alpha}$.}
%% EH - can't because of reasons stated in next paragraph

Extending this comparison to the other halos in our simulation suite
is not straightforward, because the rather large deviations from
spherical symmetry present in many cases preclude the use of the
epicyclic approximation to compute closed orbits. Our set of 11
simulations is also too small to allow for a statistically meaningful
comparison between simulated halos and LSB rotation curves.  The case
of G4 is nevertheless encouraging, as it signals that perturbations of
the right magnitude and the right shape can result in ``core-like''
rotation curves such as those observed in some LSB galaxies.

\section{Conclusions}
\label{sec:concl}

We have examined the three-dimensional shape of the gravitational
potential in seven Milky Way-sized halos and four dwarf galaxy-sized
halos simulated in the concordance \LCDM cosmology.  We use the
ellipsoid fitting method of \cite{SPRINGEL04} to measure the shape of
isopotential surfaces as a function of distance from the centre of the
halo. The isopotential surfaces are well described by concentric
ellipsoids whose principal axes remain aligned across the whole halo,
and whose minor axes tend to be parallel to the angular momentum of
the halo.

We find that the axial ratios of the isopotential surfaces decrease
sharply inwards inside the characteristic scale radius of the halo
mass profile, approaching central values of $(c/a)_0 = 0.72 \pm 0.04$
and $(b/a)_0=0.78 \pm 0.08$. The potential becomes more spherical and
less axisymmetric from the centre outwards, as $c/a$ increases more
gradually than $b/a$. Such radial dependence is well captured by a
simple fitting formula, eq.~(\ref{eq:tanhfit}), where the characteristic
radial scale for $c/a$ is about twice that of $b/a$. Incorporating the
radial variation in shape in models of CDM halos represents a
significant improvement over the analytic halo potentials generally
implemented in N-body simulations.

We use the analytic closed loop orbit solutions presented in
\cite{HAYASHI06} to investigate the kinematics of thin, filled gaseous
disks embedded in such halos. We find that the radially-dependent
ellipticity of the potential on the principal planes of the halo
generally resembles the perturbation needed to reconcile ``core-like''
rotation curves with cuspy halo profiles. The amplitude of the
ellipticity in the potential is typically larger than needed to obtain
linearly-rising long-slit rotation curves according to the treatment
presented in HN.

This is important, as the simulations do not include baryons, and the
assembly of a baryonic disk will tend to circularize the potential, by
(i) inducing changes in the actual mass distribution of the dark halo
\citep[][Abadi et al, in preparation]{DUBINSKI94,GUSTAFSSON06}, and by (ii) ``opposing'' the
halo ellipticity---the disk would itself be elliptical but rotated by
90$^o$ relative to the halo potential. The magnitude of the
corrections implied by these two effects is at present quite
uncertain, and will require simulations more precisely tailored to
reproducing the formation of a system with the observed properties of
an LSB galaxy.

Because of these uncertainties it would be premature to conclude that
halo triaxiality is the main cause of both the wide variety in LSB
rotation curve shapes and the existence of rotation curves suggestive
of the presence of constant-density cores \citep{HAYASHI04}. Still,
our results demonstrate that radially-varying ellipsoidal potentials
are expected to be the rule rather than the exception in CDM halos;
and indicate that this ``natural'' feature of CDM halo structure needs
to be included in more sophisticated analyses of LSB rotation curves
and of other observational probes of the shape of CDM halos.

\section*{Acknowledgements.} 
\label{sec:acknowledgements}

% We thank Simon White for useful discussions and Volker Springel for
% providing the shape measurement code.  
The simulations were performed as part of the programme of the Virgo
Consortium using Cosmology Machine supercomputer at the Institute for
Computational Cosmology, Durham, and the High Performance Computing
Facility at the University of Victoria.  We thank Simon White, Adrian
Jenkins, Chris Power, and Carlos Frenk for useful discussions and for
their help with the numerical simulations.  We also thank the anonymous 
referee for comments which improved the manuscript. JFN acknowledges 
support from the Natural Sciences \& Engineering Research Council of 
Canada (NSERC), the Alexander von Humboldt Foundation and the Leverhulme
Foundation.

\bibliographystyle{mn2e}
%\bibliography{/Users/ehayashi/Documents/pubs/stan}
\bibliography{../../stan}

\clearpage
\onecolumn

\landscape
\begin{table*}
\begin{center}
\caption{Main parameters of simulated halos}
\begin{tabular}{clrrrcccccccccccccc}
\hline
Label & $\epsilon$ &$N_{200}$ & $M_{200}$ & $r_{200}$ & 
$r_{\rm conv}$ & $r_s$ & \multicolumn{3}{c}{potential $c/a$} &
\multicolumn{3}{c}{potential $b/a$} & \multicolumn{3}{c}{mass $c/a$} &
\multicolumn{3}{c}{mass $b/a$} \\
      & [$h^{-1}$ kpc] &      &  [$h^{-1} \, M_{\odot}$] & [$h^{-1}$ kpc] &
[$h^{-1}$ kpc] & [$h^{-1}$ kpc] & $\alpha$ & $r_\alpha/r_s$ & $\gamma$ & $\alpha$ & $r_\alpha/r_s$ & $\gamma$ & $\alpha$ & $r_\alpha/r_s$ & $\gamma$ & $\alpha$ & $r_\alpha/r_s$ & $\gamma$ \\
\hline
D1 & $0.0625$  & $784980$  & $7.81\times 10^{9}$  & $32.3$   & $0.34$& $2.59$ &0.052&  3.60&   1.39&   0.024&  0.69&   0.82 &0.15&12.75&0.95&0.08&2.19&0.88\\
D2 & $0.0625$  & $778097$  & $9.21\times 10^{9}$  & $34.1$   & $0.37$& $2.43$ &0.058&  2.87&   1.09&   0.042&  0.73&   3.01 &0.15&17.69&1.14&0.12&1.42&1.57\\
D3 & $0.0625$  & $946421$  & $7.86\times 10^{9}$  & $32.3$   & $0.33$& $2.94$ &0.065&  3.49&   1.00&   0.060&  1.20&   1.22 &0.20&12.18&0.78&0.16&3.90&1.42\\
\smallskip						     	                                                                                            
D4 & $0.0625$  & $1002098$ & $9.72\times 10^{9}$  & $34.7$   & $0.32$& $2.06$ &0.065&  3.14&   1.54&   0.048&  2.68&   1.90 &0.16&9.09&1.34&0.12&6.44&1.57\\ 
G1 & $0.15625$ & $3447447$ & $2.29\times 10^{12}$ & $214.4$  & $1.42$& $23.2$ &0.100&  1.37&   1.27&   0.097&  0.86&   1.76 &0.20&9.08&1.04&0.19&2.30&1.56\\ 
G2 & $0.5$     & $4523986$ & $2.93\times 10^{12}$ & $232.6$  & $1.25$& $16.8$ &0.075&  7.14&   1.08&   0.066&  4.21&   0.91 &0.19&12.31&2.83&0.16&8.86&1.45\\
G3 & $0.45$    & $2661091$ & $2.24\times 10^{12}$ & $212.7$  & $1.65$& $28.0$ &0.075&  2.24&   0.81&   0.053&  0.87&   0.68 &0.17&5.04&1.25&0.13&1.56&1.27\\ 
G4 & $0.3$     & $3456221$ & $1.03\times 10^{12}$ & $164.0$  & $1.01$& $12.3$ &0.078&  0.53&   1.00&   0.062&  0.22&   1.85 &0.23&1.34&0.81&0.16&0.42&1.29\\ 
G5 & $0.35$    & $3913956$ & $1.05\times 10^{12}$ & $165.0$  & $1.02$& $13.8$ &0.059&  7.66&   0.56&   0.019&  $\gg10$&0.07 &0.38&0.60&0.77&0.24&1.34&0.93\\ 
G6 & $0.35$    & $3739913$ & $9.99\times 10^{11}$ & $162.5$  & $1.03$& $15.3$ &0.078&  0.43&   1.45&   0.066&  0.46&   1.68 &0.21&0.97&1.26&0.16&1.18&1.53\\ 
\smallskip						     	                                                                                            
G7 & $0.35$    & $3585676$ & $9.58\times 10^{11}$ & $160.3$  & $1.02$& $13.4$ &0.087&  0.92&   1.04&   0.068&  0.41&   1.46 &0.22&2.34&0.85&0.17&0.76&1.41\\ 
\hline
\label{tab:halos}
\end{tabular}
\end{center}
\end{table*}
\endlandscape

\clearpage

\clearpage

\clearpage

\clearpage

\clearpage

\clearpage

\clearpage

\clearpage

\clearpage

\clearpage

\end{document}